\PassOptionsToPackage{square,numbers,sort&compress}{natbib}

\documentclass[journal,twoside,web,10pt]{ieeecolor}

\usepackage{color}
\usepackage{bm}

\usepackage[utf8]{inputenc}

\newcommand{\cc}[1]{{#1}} 
\newcommand{\ccc}[1]{{\color{black} {#1}}} 

\def\D{\ccc{D_{\rm NMR}}}
\def\D{D}

\def\lunit{\,\mu{\rm m}}
\def\tunit{\,{\rm ms}}
\def\dunit{\,\mu{\rm m^2/ms}}

\def\e{\bm{e}}
\def\u{\bm{u}}
\def\g{\bm{g}}
\def\q{\bm{q}}

\def\n{\bm{n}}
\def\vecr{\bm{r}}

\def\0{{\bm{0}}}
\def\a{{\bm{a}}}
\def\b{{\bm{b}}}
\def\h{{\bm{h}}}

\def\E{{\mathbb E}}

\def\der{\,\mathrm{d}}

\PassOptionsToPackage{square,numbers,sort&compress}{natbib}

\usepackage{graphicx}

\usepackage{cite}
\usepackage{amsmath,amssymb,amsfonts}
\DeclareMathOperator\artanh{artanh}
\DeclareMathOperator\Tr{Tr}
\DeclareMathOperator\erfi{erfi}

\usepackage{generic}
\usepackage{epstopdf}
\def\BibTeX{{\rm B\kern-.05em{\sc i\kern-.025em b}\kern-.08em
    T\kern-.1667em\lower.7ex\hbox{E}\kern-.125emX}}
\begin{document}
\title{Probing surface-to-volume ratio of an anisotropic medium by diffusion {NMR} with general gradient encoding}

\author{Nicolas Moutal, Ivan~I.~Maximov, and Denis~S.~Grebenkov
\thanks{Copyright (c) 2019 IEEE. Personal use of this material is permitted. However, permission to use this material for any other purposes must be obtained from the IEEE by sending a request to pubs-permissions@ieee.org.}
\thanks{N.~Moutal and D.~S.~Grebenkov are with the PMC laboratory at CNRS and Ecole Polytechnique, Palaiseau, F-91128, France (e-mail: nicolas.moutal@polytechnique.edu and denis.grebenkov@polytechnique.edu)}
\thanks{I.~I.~Maximov is with the Norwegian Centre for Mental Disorders Research (NORMENT), Oslo University Hospital, Oslo, Norway
and with the Department of Psychology, University of Oslo, Oslo, Norway
(e-mail: ivan.maximov@psykologi.uio.no)}
}

\maketitle

\begin{abstract}
Since the seminal paper by Mitra \textit{et al.}, diffusion MR has been widely used in order to estimate surface-to-volume ratios. In 
the present work we generalize Mitra's formula for arbitrary diffusion encoding waveforms, including
recently developed q-space trajectory encoding sequences. We show that surface-to-volume ratio can be significantly misestimated 
using the original Mitra's formula without taking into account the applied gradient profile. 
In order to obtain more accurate estimation in anisotropic samples we propose an efficient and robust optimization algorithm to design 
diffusion gradient waveforms with prescribed features. \cc{Our results are supported by Monte Carlo simulations.}

\end{abstract}


\section{Introduction}
\label{section:intro}
\IEEEPARstart{I}{n} a seminal paper, Mitra {\it et al.} have derived the short-time asymptotic behavior of the time-dependent diffusion coefficient in restricted geometries \cite{Mitra1992a}:
\begin{equation}  \label{eq:Mitra}
\cc{D_{\rm MSD}(t)} = \ccc{D_0} \left(1 - \frac{1}{d}\frac{4}{3\sqrt{\pi}} \frac{S}{V} \sqrt{\ccc{D_0}t} + O(t)\right)\;,
\end{equation}
where $\ccc{D_0}$ is the intrinsic diffusion coefficient, $d$ is the space dimensionality, $S/V$ is the surface-to-volume ratio of the medium, and $O(t)$ means that the next term is at most of order of $t$. Mitra's formula describes the decrease of the \cc{time-dependent} diffusion coefficient due to restriction of spin-carrying molecules by the boundaries of the medium at short diffusion times.  Higher-order terms of Mitra's formula expansion were analyzed as well and provided additional information about the medium structure such as mean curvature, permeability and surface relaxation \cite{Mitra1992b,Mitra1993a,Latour1993a,Helmer1995a,Helmer1995b,Sen2004a,Grebenkov2007a}. 

\cc{The diffusion coefficient $D_{\rm MSD}(t)$ is defined as the ratio between the mean-squared displacement of the diffusing particles and time $2dt$.}
Using pulsed-gradient spin-echo (PGSE) experiments \cite{Stejskal1965a} $\cc{D_{\rm MSD}(t)}$ could be estimated from the diffusion signal attenuation \cc{if the gradient pulses were infinitely short. Despite the practical limitations on the gradient pulse duration,} this protocol was often \cc{applied} to estimate the surface-to-volume ratio of porous media \cite{Fordham1994a,Huerlimann1994a,Latour1994a,Helmer1995a,Mair1999a,Carl2007a}. However, such sequences typically require high gradients and do not take advantage of the experimental variety of gradient encoding schemes.

Mitra's formula \eqref{eq:Mitra} was extended to constant field gradient \cite{Swiet1994a} which received experimental validation in \cite{Helmer1995a}.
An extension to an arbitrary \cc{linear} gradient waveform was later derived in \cite{Grebenkov2007a}.
The particular case of oscillating gradients was considered in \cite{Novikov2011b}. It was recently experimentally demonstrated that such sequences make the estimation of $S/V$ accessible to small-gradient hardware, such as clinical scanners \cite{Lemberskiy2017a}. \cc{In these settings, one measures an effective diffusion coefficient $\der(t)$ that depends on the NMR sequence and in general can no longer be directly interpreted as a measure of mean-squared displacement. Fr{\o}lich \textit{et al.} obtained in \cite{Froehlich2008a} a general formula where $\der(t)$ is expressed in terms of the diffusion propagator at the boundary of the pore.}

In the article by Mitra {\it et al.}, the factor $1/d$ was claimed to be valid for any medium of dimensionality $d$, by extrapolating 
results obtained with a sphere ($d=3$), a cylinder ($d=2$), and a slab ($d=1$). It was pointed out in the review \cite{Grebenkov2007a} that 
an anisotropic medium should yield different $S/V$ ratios depending on the gradient orientation with respect to the medium. As the structure
of the medium is probed by diffusion, the diffusion length (typically of the order of microns for water) naturally distinguishes three 
types of anisotropy:
%
\begin{itemize}
\item 
The \emph{microscopic} anisotropy on much smaller scales than the
diffusion length (e.g., intracellular structure with submicron-sized
organelles);

\item 
The \emph{mesoscopic} anisotropy on scales comparable to the the
diffusion length (this is typically the size of pores, cells, or other
confining domains);

\item 
The \emph{macroscopic} anisotropy on much larger scales than the
diffusion length\cc{, that can be sensed over the size of an imaging voxel.}
\end{itemize}
The microscopic anisotropy is usually modeled via a non-isotropic
diffusion tensor $\mathcal{D}$
\cite{Basser1994a,Mattiello1994a,Basser1994b,Basser2002a}. 
Mesoscopic anisotropy, on the other hand, manifests itself in the shape of individual compartments or pores whereas macroscopic anisotropy is related to orientation dispersion \cc{of these compartments}. \ccc{For instance, diffusion tensor imaging typically characterizes macroscopic anisotropy via \cc{order parameter (OP)}, and microscopic anisotropy via micro-fractional anisotropy ($\mathrm{\mu}$FA) 
\cite{Lasic2014a,Szczepankiewicz2015a,Westin2016a}.} \ccc{The anisotropy of the medium is often described by the fractional anisotropy (FA) that depends on macro- and micro-anisotropy and can be expressed in terms of OP and $\rm \mu$FA \cite{Lasic2014a}}. \cc{Despite its importance, little work was devoted to mesoscopic anisotropy of confining media \ccc{\cite{Jespersen2019a}}. The purpose of this article is to show that it generally makes the time dependence of $\der(t)$ anisotropic, i.e. dependent on the relative orientation of the gradient sequence and the medium.}

Since short-time experiments deal with small diffusion length scales (a few microns for liquids), anisotropy tends to be relevant 
at the mesoscopic and macroscopic scales rather than at the microscopic one. For this reason, throughout this article we focus on 
mesoscopic and macroscopic anisotropy of the confining medium \cc{by considering a scalar diffusion coefficient in the sample (see extensions in Sec. \ref{section:extension_tensor})}. 
We extend previously obtained results to arbitrary gradient encoding schemes and obtain a generalization of Mitra's formula to 
gradient profiles that can change their amplitude in all directions. This is particularly important for the analysis of diffusion signal\cc{s} acquired by using $q$-space trajectory encoding schemes \cite{Westin2016a}, including, e.g., multiple pulsed-gradient sequences \cite{Mitra1995a, Cheng1999a, Shemesh2015a} and isotropic diffusion 
weighting \cite{Mori1995a,Wong1995a,Graaf2001a,Valette2012a,Eriksson2013a,Topgaard2013a,Topgaard2017a,deAlmeidaMartins2016a}.

The paper is organized as follows: in Sec. \ref{section:results}, we introduce some notations and present our generalization of Mitra's 
formula. Technical computations are detailed in Appendix \ref{section:computations}. The proposed formula differs from the 
classical one \eqref{eq:Mitra} by a \cc{dimensionless} factor $\eta$ which is shown \cc{to depend} on the structure of the medium and on the applied 
gradient waveform. 
In Sec. \ref{section:geometry}, we study the effect of structure, in particular, of the anisotropy of the confining \cc{domains}. 
We first consider a single domain and then evaluate the influence of orientation dispersion on the scale of a voxel. \cc{Exact computations for spheroids and perturbative computations for slightly non-spherical domains are provided in Appendix \ref{section:computation_S3}.}
Section \ref{section:waveform} is devoted to a design of gradient
waveforms and their influence on the estimated parameters.  We start
with the simpler case of linear encoding, for which we recover and
extend earlier results. In particular, we show that the diffusion
encoding waveform significantly influences the factor $\eta$, \cc{and its ignorance may}
lead to \cc{substantial} errors on the estimated $S/V$ ratio. The minimal and
maximal achievable values of $\eta$ are explained in Appendix
\ref{section:optim}.
After that, we turn to 3D gradient encoding schemes, with a focus on spherical encoding techniques. We show that typical spherical 
encoding sequences do not perfectly average out the mesoscopic anisotropy of the medium in the generalized Mitra's formula. \cc{Then} we 
present a simple algorithm to design \cc{various} \cc{3D} gradient sequences with prescribed properties \cc{that allows to perform a reliable estimation of the $S/V$ ratio. At the same time, we show in Appendix \ref{section:full_iso} that it is mathematically impossible to design a sequence that makes the time dependence of $\der(t)$ isotropic to all orders in $(\ccc{D_0}t)^{1/2}$.}
\cc{
In Sec. \ref{section:monte_carlo}, we present the results of Monte Carlo simulations demonstrating a very good agreement with our theory. Finally, Sec. \ref{section:extensions} presents several extensions of our results: study of the next order ($\ccc{D_0}t$) term, effect of microscopic anisotropy, generalization to multiple isolated compartments with different intrinsic diffusivities, shapes, etc.
}

\section{Results}
\label{section:results}

\begin{figure}[t]
\begin{center}
\includegraphics[width=0.8\linewidth]{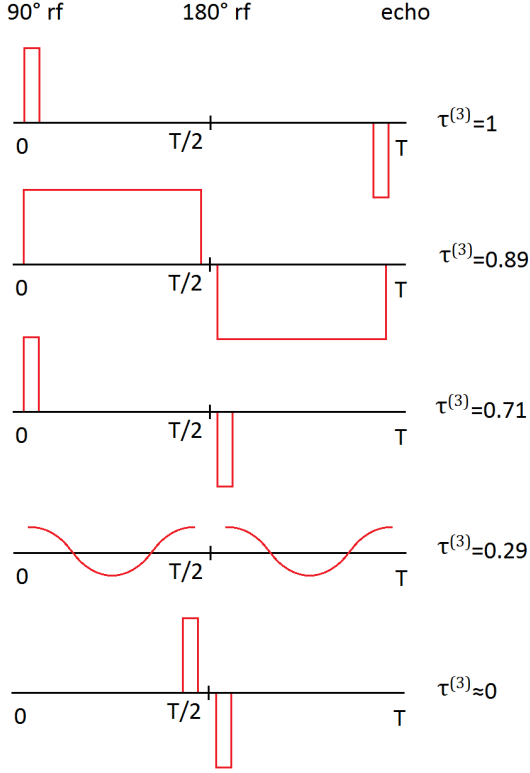}
\caption{Illustration of some gradient profiles for spin echo experiments.  We stress that these gradient profiles are ``effective'' in the 
sense that the gradients are reversed after $T/2$ to \cc{include} the effect of the $180^\circ$ rf pulse. 
The corresponding values of $\cc{\tau^{(3)}}$ introduced in Sec. \ref{section:linear} are given for each profile.
Note that $\cc{\tau^{(3)}}=1$ for the \cc{narrow pulses}-case (first panel), which corresponds to Mitra's formula \eqref{eq:Mitra} \cc{with} $t=T$.}
\label{fig:profils}
\end{center}
\end{figure}

We \cc{consider} spin-carrying molecules diffusing with \cc{scalar intrinsic} diffusivity $\ccc{D_0}$ in
a restricted domain $\Omega$, in the presence of a magnetic field
gradient $\g(t)$, with $t\in [0,T]$. Here, $t=0$ corresponds to the
beginning of the gradient sequence after the $90^\circ$
radio-frequency (rf) pulse and $t=T$ corresponds to the echo time at
which the signal is acquired (see Fig. \ref{fig:profils}).  We presume
that there are no magnetic impurities near the domain boundaries, so
that the gradient is uniform in the domain. \cc{We also assume that the intrinsic diffusivity $\ccc{D_0}$ is constant throughout the domain $\Omega$.}
\cc{
An extension to multiple isolated compartments with different intrinsic diffusivities and shapes is discussed in Sec. \ref{section:extensions_multiples}.
}
\color{black}
We define
\begin{equation}
\q(t)=\gamma\int_0^t \g(t')\der{t'}\;,
\label{eq:q_def}
\end{equation}
where $\gamma$ is the nuclear gyromagnetic ratio, and
\begin{equation}
b=\int_0^T \lvert \q(t)\rvert^2\der{t}\;
\end{equation}
is the conventional $b$-value.  The gradient profile is supposed to
obey the \cc{usual} refocusing condition 
\begin{equation} \label{eq:refocusing}
\q(T)= \cc{\gamma}\int_0^T \g(t)\der{t} = \0 .
\end{equation}
From an experimental point of view, this means that $\g(t)$ is the
``effective'' gradient which takes into account the effect of \cc{refocusing}
rf-pulses on the spins (for example, the gradient is effectively
reversed by a $180^\circ$ rf pulse)
\cite{Karlicek1980a}. This convention allows us to treat spin echo,
gradient echo, stimulated echo, and other techniques, with the same
formalism.

At small $b$-values (that is, $b\ccc{D_0} \ll 1$), the MR signal attenuation
$E$ can be written as
\begin{equation}
E\approx\exp(-b\D(T))\;,
\label{eq:S}
\end{equation}
where $\der(T)$ is the effective (or apparent) diffusion coefficient
probed by diffusion MR.  We generalize the Mitra's formula
\eqref{eq:Mitra} as
\begin{equation}  \label{eq:Mitra_gen}
\der(T) = \ccc{D_0} \left(1 - \eta \frac{4}{3\sqrt{\pi}} \frac{S}{V} \sqrt{\ccc{D_0}T} + O(T) \right)
\end{equation}
by introducing the \cc{dimensionless} prefactor $\eta$ that depends both on the
structure of the medium and on the gradient waveform.  We stress that
dependence on the waveform implies that one cannot, strictly speaking,
interpret $\der(T)$ as a measure of mean-squared displacement of randomly
diffusing molecules, except for the theoretical case of two infinitely
narrow gradient pulses.

Using mathematical methods discussed in \cc{reviews}
\cite{Axelrod2001a,Grebenkov2007a,Grebenkov2009a},
we derived in Appendix \ref{section:computations} that
\begin{equation}  \label{eq:eta}
\cc{\eta = \Tr(\mathcal{S}^{(3)}\mathcal{T}^{(3)})} \;,
\end{equation}
where \cc{$\Tr(\cdot)$} is the trace.  Here we introduced the ``temporal'' matrix
\cc{$\mathcal{T}^{(3)}$} which is a particular case of the general
\cc{$\mathcal{T}^{(m)}$} matrices,
\begin{equation}
\cc{\mathcal{T}^{(m)}=-\frac{\gamma^2 T}{2b}\int_0^T\!\!\int_0^T \!\g(t_1)\otimes \g(t_2)\left\lvert\frac{t_2-t_1}{T}\right\rvert^{m/2}\!\!\der{t_1}\der{t_2}\;,}
\label{eq:Fn}
\end{equation}
and the ``structural'' matrix
\begin{equation}
\cc{\mathcal{S}^{(3)}}=\frac{1}{S}\int_{\partial \Omega} \n \otimes \n \,\mathrm{d}S\;,
\label{eq:W}
\end{equation}
where the integration is performed \cc{over} the boundary $\partial \Omega$ of the domain $\Omega$ and $\n$ is the unit outward normal vector to the boundary.
\cc{In the above formulas, $\otimes$ is the outer product}: if $\a$ and $\b$ are vectors, then $\a \otimes \b$ is a matrix with components
\begin{equation}
(\a \otimes \b)_{ij}=a_i b_j\;.
\end{equation}
\cc{Note that $\mathcal{S}^{(3)}$ and $\mathcal{T}^{(m)}$ are tensors in the sense that under a spatial rotation or symmetry described by a matrix $\mathcal{R}$, $\mathcal{S}^{(3)}$ and $\mathcal{T}^{(m)}$ are transformed according to $\mathcal{S}^{(3)}\to \mathcal{R}\mathcal{S}^{(3)}\mathcal{R}^{-1}$ and $\mathcal{T}^{(m)}\to \mathcal{R}\mathcal{T}^{(m)}\mathcal{R}^{-1}$.}
\cc{We note that,} with these notations, \cc{$\mathcal{T}^{(2)}$} is actually the conventional $b$-matrix renormalized by the $b$-value \cite{Basser1994a,Mattiello1994a,Basser1994b} so that (see Appendix \ref{section:computations} from Eq. \eqref{eq:int_by_parts_start} to Eq. 
\eqref{eq:int_by_parts_end} for a detailed computation)
\begin{equation}
\Tr(\mathcal{T}^{(2)})=1\;.
\end{equation}

\cc{The correction factor $\eta$ in Eq. \eqref{eq:Mitra_gen} is the result of an intricate coupling between the medium structure and the gradient sequence, which is expressed through the simple mathematical relation \eqref{eq:eta} between $\mathcal{S}^{(3)}$ and $\mathcal{T}^{(3)}$.
}
\cc{Note that $\mathcal{S}^{(3)}$ and all $\mathcal{T}^{(m)}$ matrices, in particular $\mathcal{T}^{(3)}$, are dimensionless. As a consequence, $\eta$ is invariant under dilatation of the gradient waveform, dilatation of the time interval $[0,T]$ and dilatation of the domain $\Omega$.}
The higher-order terms in the asymptotic expansion \eqref{eq:Mitra_gen} involve increasing powers of $\sqrt{\ccc{D_0}T}$ associated with the 
temporal matrices \cc{$\mathcal{T}^{(m)}$} with increasing integer $m$. These matrices are coupled \cc{to} structural matrices \cc{$\mathcal{S}^{(m)}$} (such as in Eq. 
\eqref{eq:eta}) that characterize the medium structure and properties such as curvature, permeability or surface relaxation.  However, these 
properties do not affect the first-order term in \eqref{eq:Mitra_gen}, on which we focus in this paper. \cc{As an example, the second-order, $\ccc{D_0}T$, term and the associated matrix $\mathcal{S}^{(4)}$ are discussed in Sec. \ref{section:extensions_DT}.}

Mitra's formula \eqref{eq:Mitra} was  derived for PGSE experiments with (infinitely) short gradient pulses, where $t=\Delta$ is the 
inter-pulse time. We emphasize that for general gradient profiles, $\Delta$ is not defined anymore, and we use instead \cc{the} echo time $T$ in our 
generalized formula \eqref{eq:Mitra_gen}. If we compare the two formulas by setting $t=T$ (which corresponds to the profile shown on the 
first panel in Fig. \ref{fig:profils}), we see that Mitra's formula corresponds to the simple expression
\begin{equation}
\eta_{\rm Mitra} = 1/d\;.
\label{eq:eta_Mitra}
\end{equation}
\cc{Below we generalize this relation to arbitrary medium structures (Sec. \ref{section:geometry}) and gradient profiles (Sec. \ref{section:waveform}).}
%


\section{Dependence on the structure}
\label{section:geometry}
\subsection{Simple shapes}

For any bounded domain $\Omega$, the matrix $\mathcal{S}^{(3)}$ is
symmetric, positive-definite, and one has $\Tr(\mathcal{S}^{(3)})=1$.  For
example, if $\Omega$ is a sphere, one \cc{gets}
$\mathcal{S}^{(3)}_{\text{sphere}}=\mathcal{I}/3$, which is invariant under
any spatial rotations of the medium, as expected.  Throughout the article, we
call such matrices\cc{, that are proportional to the $3\times 3$ unit matrix $\mathcal{I}$,}  ``isotropic''.
%
However, the same result holds if $\Omega$ is a cube, i.e.  the cube
is also qualified as isotropic by the $\mathcal{S}^{(3)}$ matrix. Clearly,
the matrix $\mathcal{S}^{(3)}$ does not uniquely characterize the shape of
$\Omega$.

Let us now consider the example of a rectangular parallelepiped. We choose its sides along the axes $(\e_x,\e_y,\e_z)$ and denote their length\cc{s} by $a$, $b$, $c$. Then the normal vector $\n$ is either $\pm \e_x$, $\pm \e_y$, or $\pm\e_z$ depending on the facet of the parallelepiped, and by integrating over each facet we get
\begin{equation}
\cc{\mathcal{S}^{(3)}}=\frac{1}{bc+ca+ab}\begin{bmatrix}bc&0&0\\0&ca&0\\0&0&ab\end{bmatrix}\;.
\label{eq:W_parallelepiped}
\end{equation}
\cc{This simple example shows that}, by varying $a$, $b$, $c$, \cc{and applying rotations,} the matrix \cc{$\mathcal{S}^{(3)}$} can be \emph{any} symmetric  positive-definite 
matrix with unit trace.



In the limit when one side of the parallelepiped (say, $c$) tends to infinity (or is much bigger than the other two), the rectangular 
parallelepiped transforms into a cylindrical domain with a rectangular cross-section and the $\cc{\mathcal{S}^{(3)}}$ matrix becomes
\begin{equation}
\cc{\mathcal{S}^{(3)}}=\frac{1}{a+b}\begin{bmatrix}b&0&0\\0&a&0\\0&0&0\end{bmatrix}\;.
\end{equation}
Note that in the special case $a=b$ (cylindrical domain with square
cross-section), one obtains the same result as for a circular cylinder
of axis $\e_z$: $\cc{\mathcal{S}^{(3)}_{\text{cyl}}}=(\mathcal{I}-\e_z\otimes\e_z)/2$.  In
the opposite limit where $a$ and $b$ are much bigger than $c$, the
parallelepiped transforms into a slab perpendicular to $\e_z$ and the
$\cc{\mathcal{S}^{(3)}}$ matrix becomes $\cc{\mathcal{S}^{(3)}_{\text{slab}}}=\e_z\otimes
\e_z$.

One recognizes in the previous examples the factor $1/d$ of Mitra's
formula \eqref{eq:Mitra}: $1/3$ for a sphere, $1/2$ for a circular
cylinder, and $1$ for a slab.  However, even in these basic cases, the
factor $\eta$ remains affected by the gradient waveform, as discussed
in Sec. \ref{section:waveform}. \cc{In Appendix \ref{section:computation_S3}, we provide additional computations of $\mathcal{S}^{(3)}$ for slightly non-spherical domains (perturbative computation) and for spheroids (exact computation). Such domains may be more accurate models of anisotropic pores in pathological tissues such as prostate tumor \cite{Lemberskiy2017b} than cylinders.}



\subsection{The effect of orientation dispersion}
\label{section:random_medium}

Now we consider a random medium consisting of infinite circular
cylinders with random orientations and random radii. \cc{Cylinders are archetypical anisotropic domains and we choose them to illustrate in an explicit way the effect of orientation dispersion of the domains.} \cc{They can also serve as a model for alveolary ducts in lungs \cite{Yablonskiy2002a} or muscle fibers \cite{Cleveland1976a,Gelderen1994a,Oudeman2016a}.} For the sake of
simplicity, we assume the radius of each cylinder to be independent
from its orientation.  Equations \eqref{eq:S} and \eqref{eq:Mitra_gen}
describe the signal on the mesoscopic scale (one cylinder). 
Within the scope of small $b$-values ($b\ccc{D_0} \ll 1$), the macroscopic signal \cc{formed by many cylinders} is:
\begin{equation}
E\approx\left\langle \exp(-b\D(T))\right \rangle \approx \exp(-b\langle \der(T)\rangle)\;,
\label{eq:signal_average}
\end{equation}
\cc{where $\langle \cdots \rangle$ denotes the average over the voxel.}
 Coming back to Eqs. \eqref{eq:Mitra_gen} and \eqref{eq:eta}, we see 
that the average of $\der(T)$ is obtained through the average of the $\cc{\mathcal{S}^{(3)}}$ matrices of the cylinders, that we now compute.

From the previous section, the $\cc{\mathcal{S}^{(3)}}$ matrix of a cylinder oriented along any direction $\u$ (where $\u$ is a unit vector) is
\begin{equation}
\cc{\mathcal{S}^{(3)}_{\text{cyl}}}(\u)=\frac{1}{2}\left(\mathcal{I}-\u\otimes\u\right)\;.
\end{equation}
Moreover, for \cc{one} cylinder of radius $R$, one has $S/V=2/R$, thus the
voxel-averaged effective diffusion coefficient reads
\begin{equation*}
\langle \der(T) \rangle=\ccc{D_0}\left(1-\frac{4 \sqrt{\ccc{D_0}T}}{3\sqrt{\pi}}\left\langle\!\frac{2}{R}\!\right\rangle \Tr(\langle\cc{\mathcal{S}^{(3)}}\rangle \cc{\mathcal{T}}^{(3)})+O(T)\!\right).
\end{equation*}
The averaged matrix $\langle\cc{\mathcal{S}^{(3)}}\rangle$ depends on the angular distribution of the cylinder orientations. For 
example, a distribution with a rotation symmetry around the $z$-axis yields
\begin{equation}
\langle\cc{\mathcal{S}^{(3)}}\rangle
=\frac{1}{6}\begin{bmatrix} 2+\cc{p} & 0 & 0 \\ 0 & 2+\cc{p} & 0 \\ 0 & 0 & 2-2\cc{p} \end{bmatrix}\;,
\end{equation}
where $\cc{p}$ is the orientation order parameter (OP) of the medium that is defined as
\begin{equation}
\cc{p}=\langle 3\cos^2\theta - 1\rangle/2=\langle 3 u_z^2-1 \rangle/2\;,
\end{equation}
where $\theta$ is the random angle between the axis of the cylinder
$\u$ and the symmetry axis $\e_z$.  The parameter $\cc{p}$ can take any
value from $-1/2$ (for $\theta=\pi/2$, i.e. all the cylinders are in
the $x-y$ plane) to $1$ (for $\theta=0$, i.e. all the cylinders are
aligned with $\e_z$). The special value $\cc{p}=0$ corresponds to an
isotropic matrix $\cc{\mathcal{S}^{(3)}} = \mathcal{I}/3$ and can be obtained,
for example, with a uniform distribution
\cite{Lasic2014a,Szczepankiewicz2015a,Westin2016a}. 

The orientation order parameter has direct analogies with other
diffusion models describing the water diffusion in strongly
anisotropic medium. For instance, if randomly oriented \cc{fibers} \cc{obey} a Watson distribution of
parameter $\kappa$
\cite{Tariq2016a}, then one can compute \cc{\cite{Zhang2011a}}
\begin{equation}
\cc{p}=\frac{3}{2\sqrt{\pi\kappa}\,e^{-\kappa}\,\erfi(\sqrt{\kappa})}-\frac{3}{4\kappa}-\frac{1}{2}\;,
\end{equation}
where $\erfi$ is the imaginary error function.  \cc{In the limit\cc{s of} $\kappa$ going to $-\infty$, $0$, and $+\infty$, we obtain $\cc{p}=-1/2$, $0$, and $1$, respectively.}



An important consequence of the above computations is that experiments at short diffusion times and small-amplitude gradients are 
unable to distinguish the mesoscopic anisotropy (the anisotropy of each cylinder) inside a macroscopically isotropic medium 
(uniform distribution of the cylinders). Therefore, regimes with longer diffusion times or higher gradients are needed for extracting mesoscopic diffusion information \cite{Lasic2014a,Szczepankiewicz2015a,Grebenkov2018a,Jespersen2013a}.

\section{Dependence on the gradient waveform}
\label{section:waveform}

In this section we investigate the dependence of the correction factor $\eta$ (and of higher-order terms) on the gradient waveform captured via 
the $\cc{\mathcal{T}}^{(m)}$ matrices. We begin with the simpler case, the so-called linear gradient encoding, where the gradient $\g(t)$ 
has a fixed direction and each $\cc{\mathcal{T}}^{(m)}$ matrix is reduced to a scalar. We show that significant deviations from the 
classical formula \eqref{eq:Mitra} arise depending on the chosen waveform. 

Next, in Sec. \ref{section:isotropy}, we study how the correction
factor is affected in the most general case when both gradient
amplitude and direction are time dependent.  In particular, we 
show that recently invented spherical encoding sequences
\cite{Eriksson2013a,Topgaard2013a} do not provide the {\em full mixing}
effect in the sense that $\eta$ still depends on the orientation of
the (anisotropic) medium. In order to resolve this problem we describe
in Sec. \ref{section:algorithm} a simple and robust algorithm to
design diffusion gradient profiles with desired features and
constraints.


\subsection{Linear encoding}
\label{section:linear}
If we set $\g(t)=g(t)\e$, with a constant unit vector $\e$, the $\cc{\mathcal{T}}^{(m)}$ matrices become
\begin{equation}
\cc{\mathcal{T}}^{(m)}=\tau^{(m)}\, \e\otimes \e\;,
\end{equation}
with the scalar
\begin{equation}
\cc{\tau^{(m)}=-\frac{\gamma^2T}{2b}\int_0^T\int_0^T g(t_1)g(t_2)\, \left\lvert\frac{t_2-t_1}{T}\right\rvert^{m/2}\der{t_1}\der{t_2}\;.}
\label{eq:tau_m}
\end{equation}
The correction factor $\eta$ becomes
\begin{equation}
\cc{\eta=\cc{\tau^{(3)}}\left(\e^\dagger \cc{\mathcal{S}^{(3)}} \e\right)\;,}
\label{eq:eta_linear}
\end{equation}
\cc{where $\e^\dagger$ denotes the transpose of $\e$.}
By keeping the same profile $g(t)$ and only changing the
direction of the applied gradient $\e$, the factor
\cc{$\cc{\tau^{(3)}}$} is unchanged and the factor $(\e^\dagger \cc{\mathcal{S}^{(3)}}
\e)$ allows one to probe the whole $\cc{\mathcal{S}^{(3)}}$ matrix\cc{, and thus microstructural information on the domain.} For this
purpose, one can transpose standard diffusion tensor imaging
reconstruction techniques \cite{Basser1994a} to our case: by choosing
multiple non-coplanar directions for $\e$, one obtains a system of
linear equations on the coefficients of $\cc{\mathcal{S}^{(3)}}$ that can be
solved numerically. Bearing in mind that $\cc{\mathcal{S}^{(3)}}$ is symmetric
positive-definite matrix with trace one, one needs at least $6$
diffusion directions to estimate $5$ independent coefficients of the
$\cc{\mathcal{S}^{(3)}}$ matrix and the $S/V$ ratio.

For a $\cc{\mathcal{S}^{(3)}}$ matrix such as the one of a parallelepiped
in Eq. \eqref{eq:W_parallelepiped}, the factor $\eta$ takes different values
depending on the gradient direction $\e$. \cc{Note that the extremal values of $(\e^\dagger\mathcal{S}^{(3)}\e)$ are given by the minimal and maximal eigenvalue of $\mathcal{S}^{(3)}$. In other words, the relative difference between the extremal eigenvalues of $\mathcal{S}^{(3)}$ indicates the magnitude of the induced error on the estimation of $S/V$.} For instance, if one applies
the gradient in a direction perpendicular to the smallest facets of
parallelepiped, one probes the $S/V$ ratio of these facets, not of the
whole structure \cc{(see Eq. \eqref{eq:W_parallelepiped})}. Although this example is specific, the conclusion is
general: the mesoscopic anisotropy of a confining domain, captured via
the matrix $\cc{\mathcal{S}^{(3)}}$, can significantly bias the estimation of the
surface-to-volume ratio. This circumstance was ignored in some former
studies with application of the classical Mitra's formula,
which is only valid for isotropic domains.  While spherical encoding
scheme aims to resolve this issue by mixing contributions from
different directions, we will see in Sec. \ref{section:isotropy} that
this mixing is not perfect for formerly proposed spherical encoding
sequences.

In the remaining part of this subsection, we consider the particular
case of isotropic (e.g., spherical) domains with $\cc{\mathcal{S}^{(3)}}=\mathcal{I}/3$ so
that the structural aspect is fully decoupled from the temporal one.
In this case, Eq. \eqref{eq:eta} yields
\begin{equation} \label{eq:alpha}
\eta=\frac{\cc{\tau^{(3)}}}{3} \;,
\end{equation}
and we can focus on the temporal aspect (gradient waveform) captured
via the factor $\cc{\tau^{(3)}}$. \cc{Note that the original Mitra's formula corresponds to $\tau^{(3)}=1$ (see Eq. \eqref{eq:eta_Mitra}).
}


Figure \ref{fig:profils} shows \cc{several} examples of temporal profiles and the corresponding  values of $\cc{\tau^{(3)}}$. The maximum attainable value of 
$\cc{\tau^{(3)}}$ is slightly over $1$ (around $1.006$), see Appendix \ref{section:optim} for more details. \cc{Counter-intuitively, the maximal value of $\cc{\tau^{(3)}}$ is not $1$ while the profile with infinitely narrow pulses does not provide its maximum.}
%
\cc{The infimum of $\tau^{(3)}$ is $0$; in fact,}
one can achieve very small values of $\cc{\tau^{(3)}}$ by using very fast oscillating gradients. \cc{Indeed}, for sinusoidal gradient waveforms of 
angular frequency $\omega$, one has $\cc{\tau^{(3)}} \sim \omega^{-1/2}$, in the limit $\omega T \gg 1$ \cc{(see Appendix \ref{section:optim} and Refs. \cite{Novikov2011b,Lemberskiy2017a})}.

This finding has an important practical consequence: if one ignores the factor $\cc{\tau^{(3)}}$ and uses the original Mitra's formula (for 
which $\cc{\tau^{(3)}}=1$), one can significantly underestimate the surface-to-volume ratio (by a factor $1/\cc{\tau^{(3)}}$) and, thus, 
overestimate the typical size of compartments.

\subsection{Isotropy and spherical encoding}
\label{section:isotropy}

Microscopic anisotropy is usually modeled via a non-isotropic
diffusion tensor $\mathcal{D}$, and the expression \eqref{eq:S} for
the diffusion signal becomes
\begin{equation}
E\cc{\approx}\exp\left(-b\Tr\left(\mathcal{T}^{(2)} \mathcal{D}\right)\right)\;.
\label{eq:S_anis}
\end{equation}
Typical spherical encoding sequences
\cite{Mori1995a,Cheng1999a,Wong1995a,Graaf2001a,Valette2012a,Eriksson2013a,Topgaard2013a}
aim to average out the microscopic anisotropy of the medium by
applying an encoding gradient with time-changing direction.
Mathematically, the goal is to obtain an isotropic
$\cc{\mathcal{T}}^{(2)}$ matrix, $\cc{\mathcal{T}^{(2)}= \mathcal{I}/3}$, so that the
signal in Eq. \eqref{eq:S_anis} depends only on the trace
$\Tr(\mathcal{D})$ and thus yields the same result for any orientation
of the medium.  We recall that throughout the paper, we call a matrix isotropic if it
is proportional to the unit matrix $\mathcal{I}$ (in other words, its eigenvalues are equal to each other).

Mesoscopic anisotropy manifests itself in the $\cc{\mathcal{S}^{(3)}}$ matrix of
individual compartments, as we explained in
Sec. \ref{section:geometry}.
%
%
Thus, from Eq. \eqref{eq:Mitra_gen} we can deduce that mesoscopic
anisotropy is averaged out (at the order $\sqrt{\ccc{D_0}T}$) by the gradient sequence {\em only} if
$\cc{\mathcal{T}}^{(3)}$ is isotropic.
In this case, the factor $\eta$ does not depend on the orientation of the mesoscopically anisotropic medium \cc{nor on its actual shape}, and one can estimate precisely the surface-to-volume ratio of the medium. Moreover, from Eq. \eqref{eq:eta} we see that in this case, $\eta$ can be read directly from the expression of $\cc{\mathcal{T}}^{(3)}$:
\begin{equation}
\cc{\mathcal{T}}^{(3)}_{\text{iso}}=\eta \, \mathcal{I}\;.
\label{eq:F3_iso}
\end{equation}
Similarly, the isotropy condition for the matrices
$\cc{\mathcal{T}}^{(4)}, \cc{\mathcal{T}}^{(5)},\ldots$ would be needed if the
higher-order terms of \cc{expansion \eqref{eq:Mitra_gen}} were considered.
%

Hence, the natural question arises: ``Do the former spherical encoding
sequences that were designed to get an isotropic $\cc{\mathcal{T}}^{(2)}$
(or $b$) matrix \cite{Westin2016a}, produce isotropic
$\cc{\mathcal{T}}^{(m)}$ matrices (or at least $\cc{\mathcal{T}}^{(3)}$)?''.
%
%
%
%
For instance, for the q-Space Magic-Angle-Spinning (q-MAS) sequence
\cite{Eriksson2013a,Topgaard2013a} we obtain
\begin{equation}
\cc{\mathcal{T}}^{(3)} =\begin{bmatrix}0.14&0&0\\0&0.28&0.10\\0&0.10&0.17\end{bmatrix} \;.
\label{eq:T3_top}
\end{equation}
This matrix has eigenvalues $0.11, 0.14, 0.33$ and is thus not
isotropic.
%
\color{black}
\cc{Similarly, a triple diffusion encoding (TDE) sequence \cite{Mori1995a} (where three identical PGSE sequences are applied along three orthogonal direction in space) yields
\begin{equation}
\cc{\mathcal{T}}^{(3)} =\begin{bmatrix}0.19&0.08&0.05\\0.08&0.19&0.08\\0.05&0.08&0.19\end{bmatrix} \;,
\label{eq:T3_TDE}
\end{equation}
with eigenvalues $0.10, 0.14, 0.34$. Note that, although the diagonal elements of the matrix are identical, it is not isotropic because of the off-diagonal elements. The above matrix corresponds to a TDE sequence where each PGSE sequence is made of infinitely narrow pulses  with spacing $\Delta=T/3$.
}
\color{black}
\cc{One could also consider} the FAMEDcos sequence \cite{Vellmer2017a}, for which we get
\begin{equation}
\cc{\mathcal{T}}^{(3)}=\begin{bmatrix}0.13&0&0.012\\0&0.11&0\\0.012&0&0.10\end{bmatrix}\;,
\end{equation}
which is also not isotropic \cc{(with eigenvalues: $0.09, 0.11, 0.13$).}
All spherical encoding schemes that we could find in the literature
produce anisotropic $\cc{\mathcal{T}}^{(3)}$ matrices.

\cc{In order to illustrate the errors induced by such sequences in the estimation of the surface-to-volume ratio, let us apply the q-MAS sequence for the case of an infinite circular cylinder.
We denote by $(\e_1,\e_2,\e_3)$ the orthogonal basis of eigenvectors and by $(\lambda_1,\lambda_2,\lambda_3)=(0.11,0.14,0.33)$ the corresponding eigenvalues of the $\cc{\mathcal{T}}^{(3)}$ matrix in Eq. \eqref{eq:T3_top} (see Fig. \ref{fig:sequence_top} for the orientation of these axes with respect to the q-space plot of the sequence).
If the cylinder is oriented along $\e_3$, one obtains $\eta=(\lambda_1+\lambda_2)/2=0.13$. However, if the cylinder is oriented along $\e_1$, then $\eta=(\lambda_2+\lambda_3)/2=0.24$, which is nearly twice as large. In other words, the estimated $S/V$ ratio is twice as large in the second situation than in the first one. This artifact is a direct consequence of the differences between the eigenvalues of the $\cc{\mathcal{T}}^{(3)}$ matrix, i.e. its anisotropy.
\begin{figure}
\centering
\includegraphics[width=0.6\linewidth]{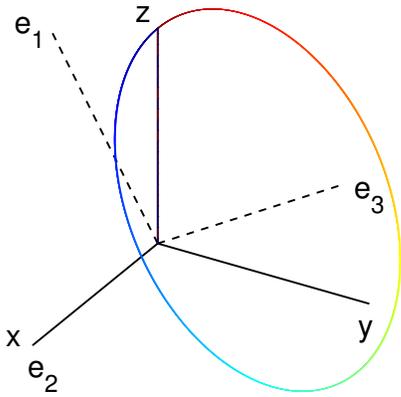}
\caption{\cc{Plot of $\q(t)$ for the q-MAS sequence. The color encoding of the trajectory represents time, from $t=0$ (blue) to $t=T$ (red). The additional axes are directed along the eigenvectors ($\e_1,\e_2,\e_3)$ of the $\mathcal{T}^{(3)}$ matrix \eqref{eq:T3_top} of the sequence.}}
\label{fig:sequence_top}
\end{figure}
}

\color{black}
\subsection{How to obtain isotropic matrices?}
\label{section:algorithm}

The question in the subsection title can be restated in an algebraic
language: how to find three functions $g_x(t), g_y(t), g_z(t)$ with
zero mean (see Eq. \eqref{eq:refocusing}) that are mutually
``orthogonal'' and have the same ``norm'' for a given set of symmetric
bilinear forms $\ccc{\varphi}_m, m=2,3,\ldots$, with
\begin{equation}
\cc{ 
\ccc{\varphi}_m(f_1,f_2)=-\frac{\gamma^2T}{2b}\!\!\int_0^T\!\!\!\int_0^T\!\! f_1(t_1)f_2(t_2)\left\lvert\frac{t_1-t_2}{T}\right\rvert^{m/2}\!\!\der{t_1}\der{t_2}\;.
}
\label{eq:phi_norm}
\end{equation} 
Since the space of functions with zero mean is infinite-dimensional,
we can be confident in finding such three functions.  However,
Eq. \eqref{eq:phi_norm} involves a non-integer power of time that
prevents us from getting analytical solution for this problem. Thus,
we design a simple algorithm for generating the gradient sequences
that satisfy these conditions.

The idea is to choose a family of functions $(f_1,f_2,\ldots,f_k)$
(for example, sines or polynomials, possibly with phase jumps at
$T/2$) and to search for $g_x(t),g_y(t),g_z(t)$ as linear combinations
of the basis functions. This is a generalization of the classical sine
and cosine decomposition which was already used in the context of
waveform optimization
\cite{Topgaard2013a}.
Mathematically, this means that
\begin{equation}
\begin{pmatrix} g_x(t)\\g_y(t)\\g_z(t)\end{pmatrix}=X\begin{pmatrix}f_1(t)\\f_2(t)\\\vdots\\f_k(t)\end{pmatrix}\;,
\end{equation}
where $X$ is a $3\times k$ matrix of coefficients to be found. Now we
define the $k\times k$ matrices $\Phi^{(m)}$ by
\begin{equation}
\cc{\Phi}^{(m)}_{i,j}=\ccc{\varphi}_m(f_i,f_j)\;,\quad m=2,3,\ldots
\end{equation}
In this way, one can compute directly the $\cc{\mathcal{T}}^{(m)}$ matrices according to
\begin{equation}
\cc{\mathcal{T}}^{(m)}=X\cc{\Phi}^{(m)}X^\dagger \;.
\label{eq:F_n_XAX}
\end{equation}
The problem is then reduced to an optimization problem for the matrix $X$, which can be easily done numerically.  In other words, one searches for {\em a} matrix $X$ that ensures
the isotropy of the matri\cc{x} $\cc{\mathcal{T}}^{(3)}$. \cc{In the same way, one can generate a sequence with both isotropic $\mathcal{T}^{(2)}$ and $\mathcal{T}^{(3)}$ matrices, or any other combination of isotropic $\mathcal{T}^{(m)}$ matrices.}
\cc{At the same time, we prove in Appendix \ref{section:full_iso} that there is no gradient sequence that produces isotropic $\mathcal{T}^{(m)}$ matrices simultaneously for all integers $m \geq 2$.
}

The optimization algorithm can include various additional
constraints. On one hand, one has a freedom to choose an appropriate
family $(f_1,f_2,\ldots,f_k)$, for example, to ensure smoothness of
the resulting gradient profile. Similarly, the refocusing condition
can be achieved by choosing zero-mean functions.  On the other hand,
it is also possible to add some constraints as a part of the
optimization problem. This is especially easy if the constraints can
be expressed as linear or bilinear forms of the gradient profile
$\g(t)$. For instance, each $\cc{\mathcal{T}}^{(m)}$ matrix in
\eqref{eq:Fn} is a bilinear form of the gradient profile allowing one
to express them as the simple matrix multiplication
\eqref{eq:F_n_XAX}. Another example of additional conditions consists
in imposing zeros to the designed gradient profiles. \cc{Indeed,} for practical
reasons, it is often easier to manipulate with the gradients that
satisfy to
\begin{equation}
\g(0)=\g(T/2)=\g(T)=\0\;.
\label{eq:condition_zeros}
\end{equation}
This is a linear condition on the gradient profile. If one denotes by $\mathcal{V}$ the $k\times 3$ matrix
\begin{equation}
\mathcal{V}=\begin{bmatrix}
f_1(0) & f_1(T/2) & f_1(T)\\
f_2(0) & f_2(T/2) & f_2(T)\\
\vdots &  \vdots  & \vdots \\
f_k(0) & f_k(T/2) & f_k(T)
\end{bmatrix}\;,
\end{equation}
then Eq. \eqref{eq:condition_zeros} becomes
\begin{equation}
X\mathcal{V}=\begin{bmatrix}0&0&0\\0&0&0\\0&0&0\end{bmatrix}\;.
\end{equation}
In the following, we impose the above condition to produce our
gradient waveforms.

It is worth to note that one can also generate flow-compensated
gradients, or more generally, motion artifacts suppression
techniques, \cc{by imposing} linear conditions on the gradient profile
\begin{equation}
\int_0^T t^p\g(t)\der{t} = \0\;,\quad p=1,2,\ldots,\cc{P}\;,
\label{eq:MAST}
\end{equation}
where $p=1$ corresponds to the flow compensation, and higher values of $p$ account for acceleration, pulsatility, etc. 
\cite{Pattany1987a,Laun2015a}. This condition can be rewritten in the matrix form $X\mathcal{M} =0$, where the $k\times \cc{P}$ matrix 
$\mathcal{M}$ is defined by
\begin{equation}
\mathcal{M}_{i,p}=\int_0^T t^p f_i(t)\der{t}\;,\quad p=1,2,\ldots,\cc{P}\;.
\end{equation}

Another type of optimizaton constraints can be based on hardware
limitations such as a need to minimize heat generation during the
sequence execution which amounts to minimizing the following quantity
\begin{equation}
\langle \g, \g \rangle=\int_0^T \lvert \g(t) \rvert^2\der{t}\;,
\label{eq:heat}
\end{equation}
which is a bilinear form of the gradient profile.  Similar to
representation \eqref{eq:F_n_XAX} for $\cc{\mathcal{T}}^{(3)}$, one can
define a matrix $\mathcal{H}_{i,j}=\langle f_i,f_j \rangle$ to write
Eq.\,\eqref{eq:heat} as $\langle \g, \g \rangle=\Tr\left(X \mathcal{H}
X^T\right)$, and then to include it into the optimization procedure.

The previous examples showed how linear and bilinear forms of the gradient profile can be simply expressed in terms of the 
weights matrix $X$, which allows one to perform very fast computations. The matrix corresponding to each condition (for example, 
$\cc{\Phi^{(3)}}$, $\mathcal{V}$, $\mathcal{H}$) has to be computed only once, then optimization is reduced to matrix multiplications. The 
size of the matrices involved in the computations is defined by the size of the chosen set of functions $(f_1,\ldots,f_k)$. 
Note that the set size is independent of the numerical sampling of the time interval $[0,T]$ that controls accuracy of 
the computations.

Some properties of the designed gradients do not fall into the
category of aforementioned linear or bilinear forms, e.g., ``max''
amplitude-function (i.e., one cannot impose the maximal gradient
constraint in this way). They can be included in the optimization, however one cannot apply the previous techniques in order to speed up the computation.


We have to emphasize that the ``optimal'' solution is not unique and it depends on the choice of the set $f_1,\ldots,f_k$. 
Moreover, if the set is sufficiently large and the number of degrees of freedom is greater than the number of constraints, 
then the algorithm will likely yield different solutions depending on the initial \cc{choice of} $X$ \cc{for an iterative solver}. This property can be advantageous 
in practice, as one can design many optimal solutions. The described optimization algorithm was implemented in Matlab 
(The MathWorks, Natick, MA USA) and is available upon request. It concatenates all the chosen constraints in a single \cc{matrix}-valued function $f(X)$ of the weight matrix $X$, in such a way that the constraints are expressed by the condition $f(X)=0$. This equation is then solved numerically with the Levenberg-Marquardt algorithm.

\begin{figure}[htpb]
\begin{center}
\includegraphics[width=0.49\linewidth]{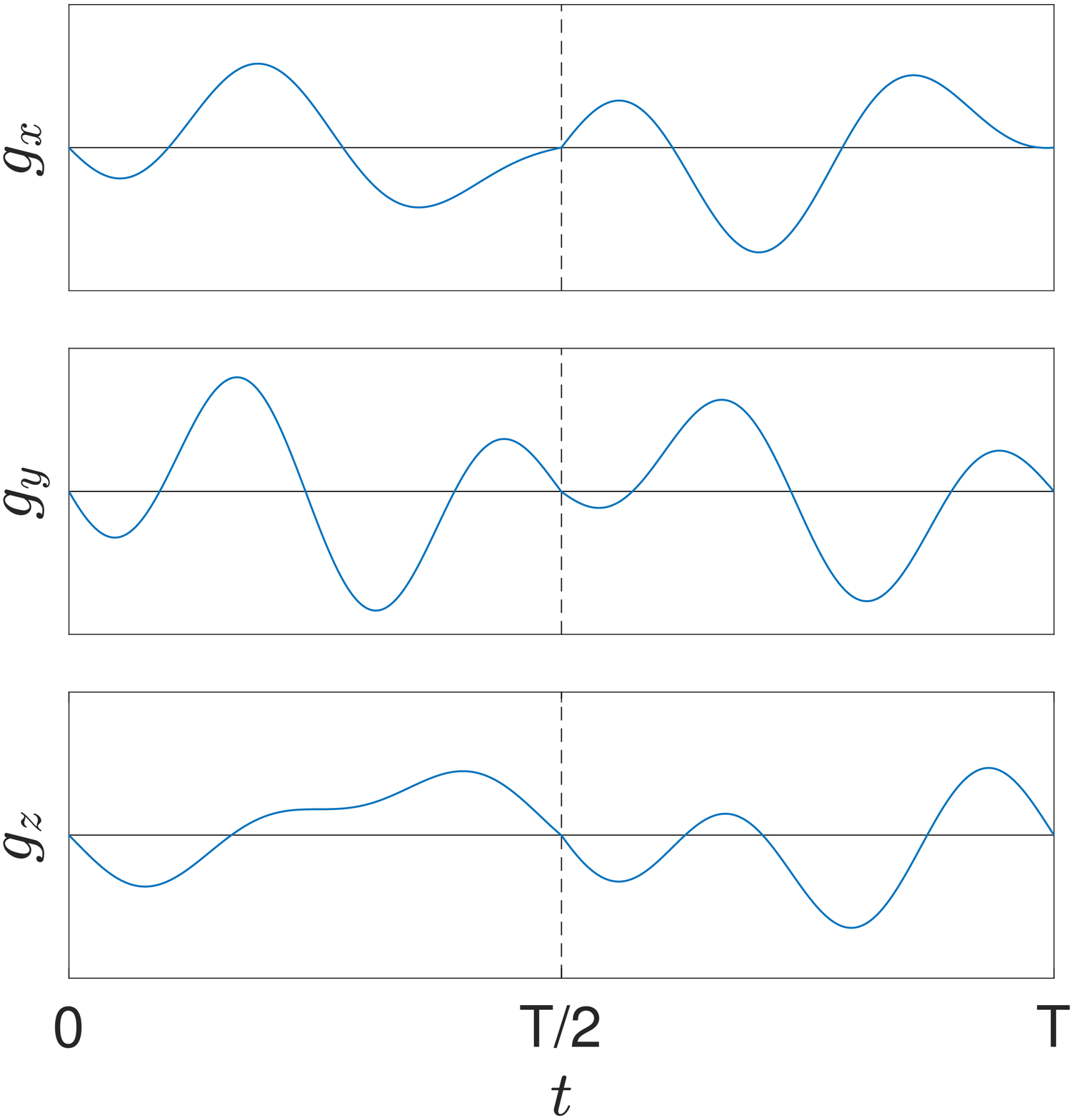}
\includegraphics[width=0.49\linewidth]{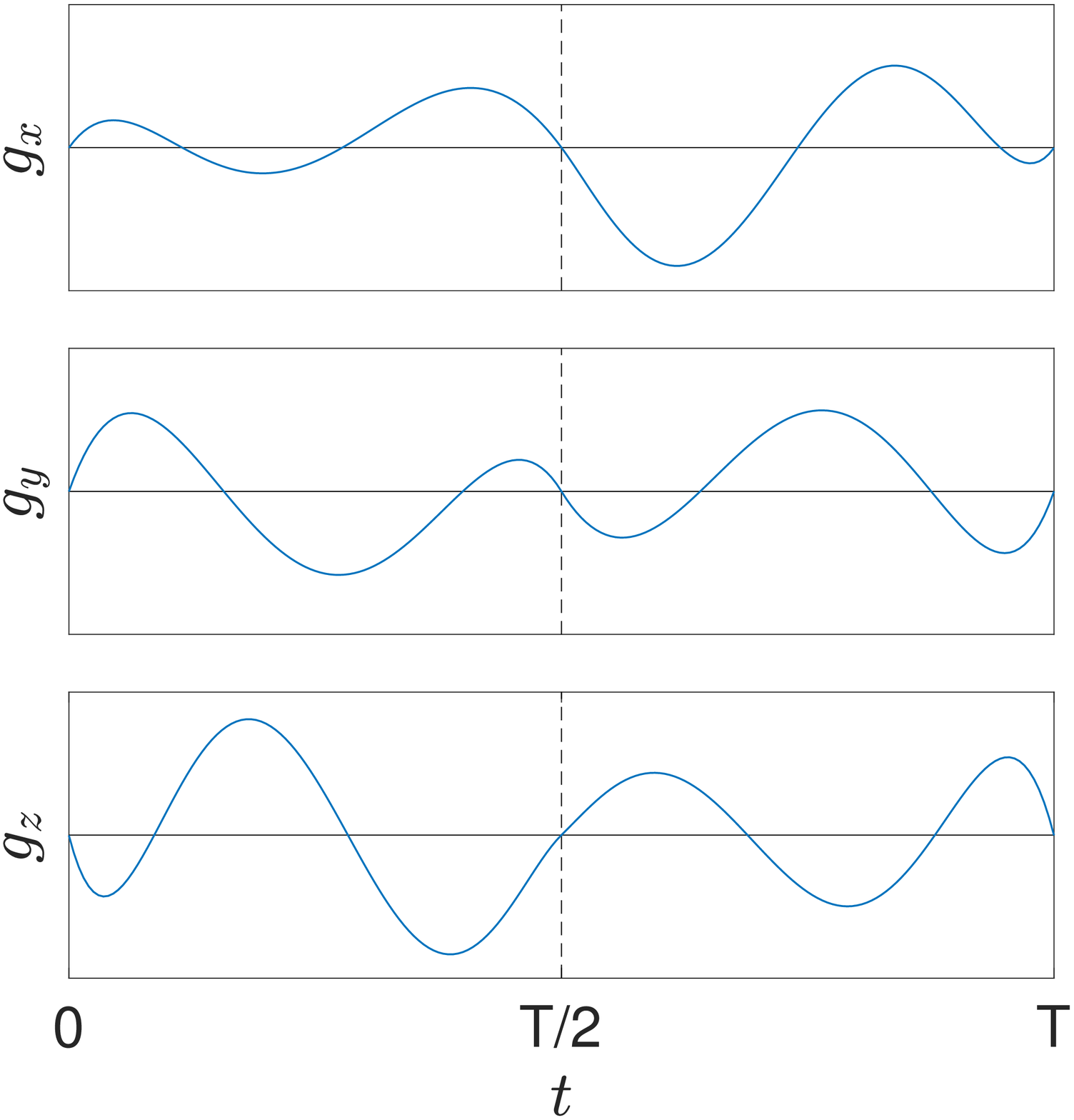}\\
\includegraphics[width=0.4\linewidth]{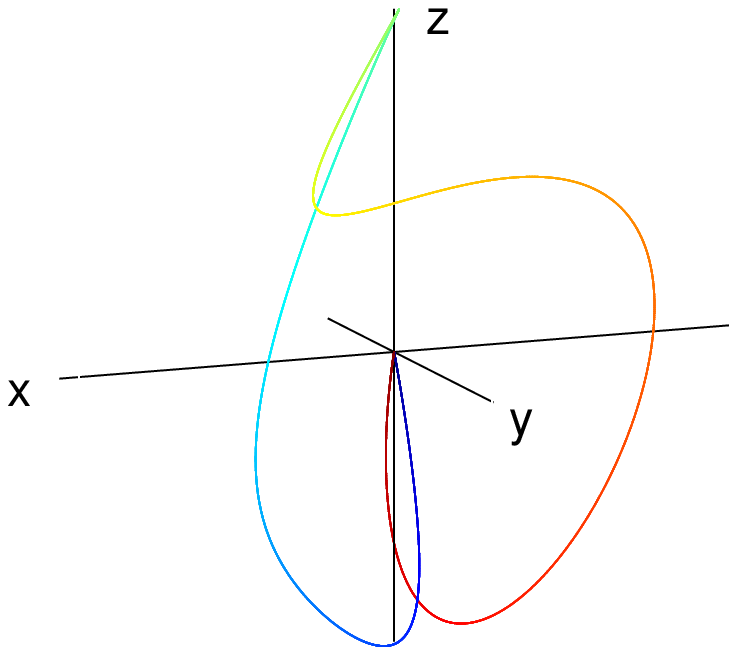}\hskip 0.1\linewidth
\includegraphics[width=0.4\linewidth]{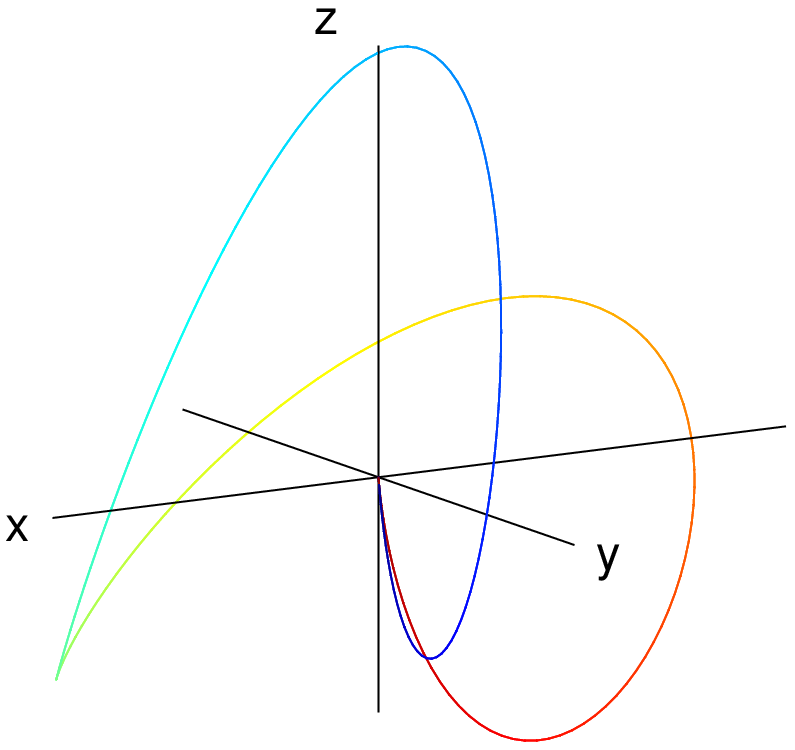}
\caption{Two examples of gradient waveforms that produce \cc{an isotropic $\cc{\mathcal{T}}^{(3)}$ matrix} and that satisfy 
Eq. \eqref{eq:condition_zeros}. Note that the gradients are ``effective'' gradients in the sense that we reversed them after the $180^\circ$ 
rf pulse at $T/2$. The bottom figure shows the corresponding $\q(t)$. \cc{The color encoding of the trajectory represents time, from $t=0$ (blue) to $t=T$ (red).} (left) the profiles are combination of $9$ piecewise sine and cosine functions 
with frequencies up to $6/T$\cc{, and in addition they satisfy isotropy of $\mathcal{T}^{(2)}$}; (right) the profiles are piecewise polynomials of order $5$,\cc{ and they satisfy
$\cc{\mathcal{T}}^{(4)}=0$.}}
\label{fig:gradient_1}
\end{center}
\end{figure}

Figure \ref{fig:gradient_1} shows two examples of gradient waveforms
that produce \cc{an isotropic $\cc{\mathcal{T}}^{(3)}$ matrix}. These profiles were obtained from \cc{two sets} with $k = 9$ functions. The first set was composed of $\cos(\pi j t/T)$ with $j=1,\ldots,5$; $\sin(\pi j t/T)$ with $j=2,4,6$; and $\varepsilon(t)\sin(4\pi t/T)$ where $\epsilon(t)$ is a piecewise constant function that is equal to $1$ on $[0,T/2]$ and $-1$ on $(T/2,T]$. \cc{We also imposed the condition of isotropy of $\mathcal{T}^{(2)}$.}
The second set was composed of a mixture of monomials, symmetrized odd monomials and antisymmetrized even monomials, with zero mean:
$(t-T/2)$, $(t-T/2)^2-T^2/12$, $(t-T/2)\lvert t-T/2 \rvert$, $(t-T/2)^3$, $\lvert t-T/2\rvert^3-T^3/32$, $(t-T/2)^4-T^4/80$, $(t-T/2)^3\lvert t-T/2\rvert$, $(t-T/2)^5$, $\lvert t-T/2 \rvert^5-T^5/192$. \cc{In this case, we imposed the condition of vanishing $\mathcal{T}^{(4)}$.}
Although the combination of symmetric and antisymmetric functions helped us to increase the number of basis functions while keeping low degree monomials or slowly oscillating sines, one could alternatively use just monomials, polynomials, or other basis functions as well. \cc{Note that there is no need to impose the orthogonality of the basis functions $f_1,\ldots f_k$.}

\begin{figure*}[h!]
\centering
\includegraphics[width=0.45\linewidth]{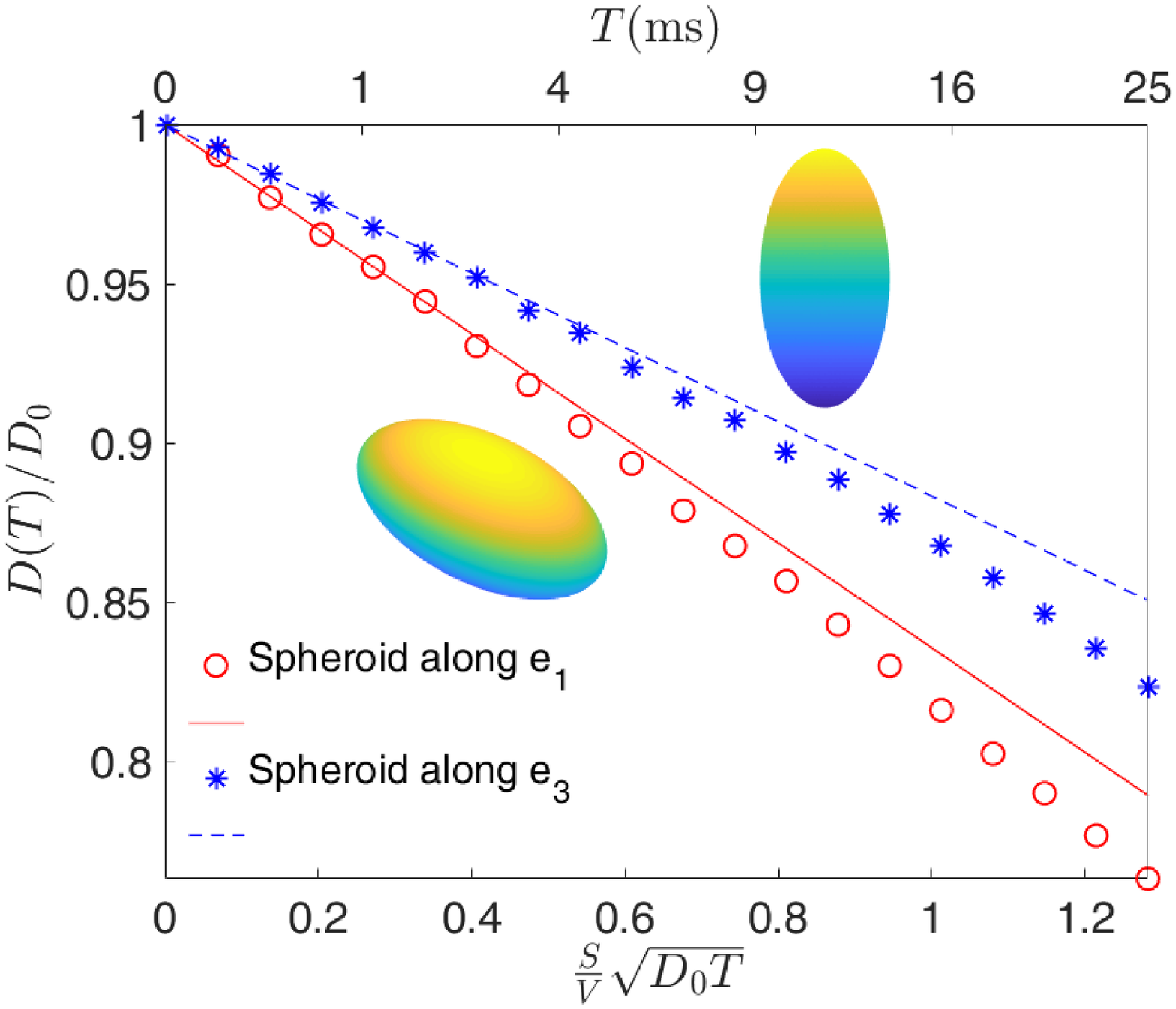}
\includegraphics[width=0.45\linewidth]{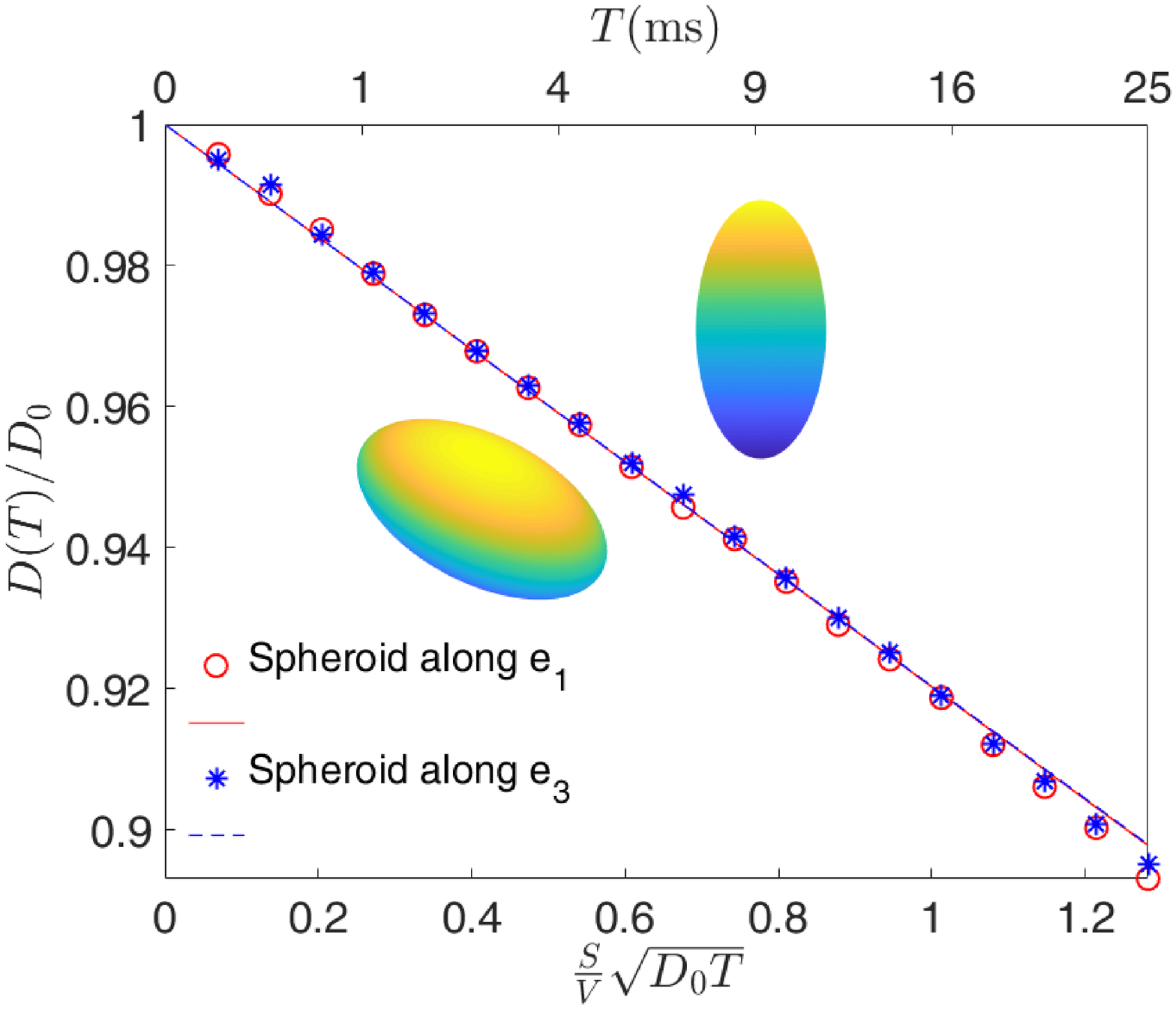}
\caption{\cc{Effective diffusion coefficient $\der(T)/\ccc{D_0}$ plotted against $\frac{S}{V}\sqrt{\ccc{D_0}T}$ inside a prolate spheroid with semi-axes $10\lunit$ and $5\lunit$ for two gradient sequences and two orientations of the spheroid. The intrinsic diffusivity is $\ccc{D_0}=1\dunit$. The simulation results are shown as symbols and the generalized Mitra formula is plotted as line. (left) q-MAS sequence: different orientations of the domain produce different $\der(T)$ curves. (right) Optimized sequence with isotropic $\mathcal{T}^{(3)}$ and zero $\mathcal{T}^{(4)}$: the $\der(T)$ curves are the same for different orientations of the domain because of the condition on $\mathcal{T}^{(3)}$. Moreover, the condition on $\mathcal{T}^{(4)}$ extends the range of validity of the theoretical formula to about $20 \tunit$.}}
\label{fig:Diff_top_iso}
\end{figure*}

\cc{Let us consider the waveform obtained in the left panel of Fig. \ref{fig:gradient_1}.} The condition of isotropy for both matrices $\cc{\mathcal{T}}^{(2)}$ and
$\cc{\mathcal{T}}^{(3)}$ yields $5+5$ equations on the components of matrix
$X$. Besides of matrices $\cc{\mathcal{T}}^{(2,3)}$,
condition \eqref{eq:condition_zeros} adds another $9$ equations on the
components of $X$. Moreover, we imposed the $b$-value so that the
algorithm satisfied $20$ conditions with $3k = 27$ degrees of freedom.

\cc{T}he gradient waveform corresponds to $\eta \approx 0.1$ \cc{and the dimensionless $b$-value is $b/(\gamma^2 G_{\rm max}^2 T^3) \approx 0.006$ (with $G_{\rm max}$ being the maximum gradient amplitude). Hence the $b$-value is about \cc{three} times smaller than what one can achieve with only the condition on the isotropy of $\cc{\mathcal{T}}^{(2)}$ \cite{Topgaard2013a}.} 
\cc{Instead of only constraining $\cc{\mathcal{T}}^{(3)}$ to be isotropic, one can in addition impose a precise value of $\eta$ by using Eq. \eqref{eq:F3_iso}. However we observed that the algorithm could not produce gradient waveforms with arbitrary values of $\eta$: there were bounds for $\eta$-values outside of which the optimization process did not converge.}
\color{black}
This behavior was expected, because even in the \cc{linear encoding} case, there were mathematical limitations for the parameter $\eta$ (see Sec. \ref{section:linear} and Appendix \ref{section:optim}).
These bounds can be extended by adding more basis functions (i.e.,
by increasing the size $k$ of their set).  Another way to extend the
bounds is to reduce the number of constraints, for example, by
dropping out the condition of isotropic $\cc{\mathcal{T}}^{(2)}$ matrix and
only keeping the condition on $\cc{\mathcal{T}}^{(3)}$. \cc{Indeed, the isotropy of $\mathcal{T}^{(2)}$ is only required in the case of a microscopically anisotropic medium, which we did not assume here (see Appendix \ref{section:extension_tensor}).
}

Interestingly, the $\cc{\mathcal{T}}^{(4)}$ matrix presents a special
case: integrating by parts in Eq.\,\eqref{eq:Fn} one can show that
\begin{equation}
\cc{\mathcal{T}}^{(4)}=\left(\int_0^T \q(t)\der{t}\right)\otimes\left(\int_0^T \q(t)\der{t}\right)\;.
\label{eq:T4}
\end{equation}
This implies that the matrix has rank one, so it cannot be proportional to the unit matrix unless it is null, that occurs under the simple condition
\begin{equation}
\int_0^T \q(t)\, \mathrm{d}t = \mathbf{0}\;.
\label{eq:condition_f4}
\end{equation}
This condition can be easily included in our optimization
algorithm. This is the case for the designed profile shown on the
right panel in Fig. \ref{fig:gradient_1}. As a consequence,
%
the corresponding term (of the order of $\ccc{D_0}T$) vanishes in the
expansion \eqref{eq:Mitra_gen}. \cc{Note that in the presence of permeation or surface relaxation, an additional term of the order of $\ccc{D_0}T$ would be present in the expansion \eqref{eq:Mitra_gen} and would not necessarily vanish if $\cc{\mathcal{T}}^{(4)}=0$. We do not consider these effects in the paper.}

\cc{The property of vanishing $\ccc{D_0}T$-order term} is well-known for cosine-based waveforms with an integer
number of periods \cite{Laun2017a}, and, indeed, such functions
automatically satisfy to Eq. \eqref{eq:condition_f4}. However, this
property is not exclusive to \cc{co}sine functions (for example, the right
panel of Fig. \ref{fig:gradient_1} was obtained with polynomial
functions). It is also easy to show that Eq. \eqref{eq:condition_f4}
is equivalent to condition \eqref{eq:MAST} for $p=1$. In other words,
flow-compensated gradient profiles automatically cancel the $\ccc{D_0}T$-order
correction term in the generalized Mitra's expansion, as it was pointed out earlier in
\cite{Laun2015a}.

\section{Monte Carlo simulations}
\label{section:monte_carlo}

\cc{
We performed Monte Carlo simulations to illustrate our theoretical results. The confining domain $\Omega$ is a prolate spheroid with major and minor semi-axes equal to $10\lunit$ and $5\lunit$. The intrinsic diffusion coefficient $\ccc{D_0}$ is $1\dunit$ and the echo time $T$ ranges from $0$ to $25\tunit$. Reflecting conditions were implemented at the boundary of the domain and the interval $[0,T]$ was divided into $200$ time steps of equal duration.}
\cc{For each value of $T$, we generated about $5\cdot 10^6$ trajectories, applied the gradient sequence and computed the effective diffusion coefficient $\der(T)$ from the second moment of the random dephasing $\phi$ of the particles: $\der(T)=\langle \phi^2 \rangle /(2b)$. In order to generate random initial positions for the particles inside the spheroid, we generated random positions inside a larger parallelepiped then discarded the particles that were outside the spheroid. We checked that the randomness in the effective number of particles was very small relatively to the number of particles (less than $0.1\%$).
}

\cc{
We chose two different gradient sequences: the q-MAS sequence \cite{Eriksson2013a,Topgaard2013a} and an optimized sequence with isotropic $\mathcal{T}^{(3)}$ and zero $\mathcal{T}^{(4)}$ such as the one in the right panel of Fig. \ref{fig:gradient_1}. 
Note that we could have replaced the q-MAS sequence by any other 3D gradient sequence from the present literature, such as triple diffusion encoding (TDE) \cite{Mori1995a}.
For each sequence, we chose two different orientations of the spheroid that yielded maximal and minimal value of $\eta$. This can be done by finding numerically the eigenvectors $(\e_1,\e_2,\e_3)$ of the $\mathcal{T}^{(3)}$ matrix (sorted by increasing eigenvalue) and then orienting the spheroid along $\e_1$ and $\e_3$, respectively (see for example Fig. \ref{fig:sequence_top}). The $\der(T)$ curves are presented on Fig. \ref{fig:Diff_top_iso}. The $\mathcal{S}^{(3)}$ matrix of a spheroid can be computed exactly (see Appendix \ref{section:computation_S3}) and we plotted simulation results alongside analytical results.
}

\cc{
The comparison between the two graphs reveals several important features. First, as we argued in the previous section, the q-MAS sequence is not isotropic with respect to mesoscopic anisotropy studied with short-time experiments. Different orientations of the spheroid yield different values of $\eta$ ($0.15$ and $0.22$, respectively) and thus, different $\der(T)$ curves. In turn, if one does not know \emph{a priori} what is the orientation of the spheroid, then it is impossible to recover its $S/V$ ratio from one $\der(T)$ curve, as $\eta$ depends on this orientation. In this case, one may estimate $\eta$ from its average over different orientations of the domain: $\eta\approx\Tr\left(\mathcal{T}^{(3)}\right)/3$. For the q-MAS sequence this would yield $\eta\approx 0.20$.
}

\cc{
On the other hand, sequences with isotropic $\mathcal{T}^{(3)}$ produce the same coefficient $\eta$ independently of the shape or orientation of the domain. Thus, one obtains the same $\der(T)$ curve for the two orientations of the spheroid that allows one to recover its $S/V$ ratio from a single measurement.
}

\cc{
Another important point lies in the range of validity of the first-order generalized Mitra formula \eqref{eq:Mitra_gen}. One can clearly see the effect of zero $\mathcal{T}^{(4)}$ matrix that extends the range of validity from about $5\tunit$ to about $20\tunit$. This comes at the price of a lower $\eta$ (here, $\eta=0.11$), meaning a slower decay of $\der(T)$, which is however compensated by the extension of the range of $T$. Note that this extension of the range of $T$ may also compensate for a smaller $b$-value.
In all these cases, the $\eta$ values are significantly different from $1/3$ given by Mitra's original formula (see Eq. \eqref{eq:eta_Mitra}).
}

\section{Extensions}
\label{section:extensions}
\cc{
In this section we examine several extensions of our results. First we investigate in more details the next order, $\ccc{D_0}T$, term of expansion \eqref{eq:Mitra_gen}. Then we turn to the case where the medium is microscopically anisotropic, i.e. the diffusivity is a tensor $\mathcal{D}$. Finally we discuss the effects of multiple compartments with different pore shapes and/or intrinsic diffusivities $\ccc{D_0}$.
}
\color{black}
\subsection{Order $\ccc{D_0}T$ term}
\label{section:extensions_DT}

From the short-time expansion of heat kernels \cite{Davies1989a,Gilkey2003a,Desjardins1994a} one can compute the next-order term of $\der(T)$ as
\begin{equation}
\frac{\der(T)}{\ccc{D_0}}=\!1-\eta \frac{4}{3\sqrt{\pi}}\frac{S}{V}\sqrt{\ccc{D_0}T}-\eta^{(4)} \frac{C_0 S}{2V}\ccc{D_0}T+O(T^{3/2}),
\end{equation}
where $\eta^{(4)}$ is a dimensionless parameter defined as
\begin{equation}
\eta^{(4)}=\Tr\left(\mathcal{S}^{(4)}\mathcal{T}^{(4)}\right)\;.
\end{equation}
In the above formula, the structural matrix $\mathcal{S}^{(4)}$ is
\begin{equation}
\mathcal{S}^{(4)}=\frac{1}{C_0 S}\int_{\partial\Omega} C \, \n\otimes\n \der{S}\;,
\end{equation}
where $C$ is the local mean curvature of the surface, i.e. $C=(R_1^{-1}+R_2^{-1})/2$, where $R_1$ and $R_2$ are the local principal radii of curvature of the boundary $\partial \Omega$ of the domain. The integral is normalized by $S$ and by the average curvature of $\partial \Omega$: $C_0= (1/S)\int_{\partial\Omega} C\d{S}$. Note that this normalization ensures that the matrix $\mathcal{S}^{(4)}$ has unit trace.

\cc{
Thus, one can potentially probe the curvature of the boundary of the domain by measuring the $\ccc{D_0}T$ correction term in the short-time expansion of $\der(T)$. Note that, as we mentioned in Sec. \ref{section:algorithm}, the $\mathcal{T}^{(4)}$ matrix has rank one so that one would need at least three measurements (for example, the same linear gradient sequence in three orthogonal directions) in order to average out the anisotropy of $\mathcal{S}^{(4)}$ and recover $C_0$. We recall that we ignore permeation and surface relaxation that manifest in the $\ccc{D_0}T$ term as well.
}

\subsection{Tensor diffusivity}
\label{section:extension_tensor}

\cc{
In this work, we specifically focused on mesoscopic anisotropy and excluded the effect of microscopic anisotropy by choosing a scalar diffusivity $\ccc{D_0}$. However, some of our results may be extended to a tensor diffusivity $\mathcal{D}$. Let us assume that the eigenvectors of $\mathcal{D}$ are directed along $\e_x$, $\e_y$, $\e_z$, with $\mathcal{D}_{xx}$, $\mathcal{D}_{yy}$, $\mathcal{D}_{zz}$ being the corresponding eigenvalues. The mean diffusivity is $\ccc{D_0}=\Tr(\mathcal{D})/3$. Let us denote by $\mathcal{S}^{(2)}$ the matrix defined by $\mathcal{S}^{(2)}=\mathcal{D}/\ccc{D_0}$. 
}

\cc{
By applying the affine mapping of matrix $\mathcal{L}=\sqrt{\mathcal{S}^{(2)}}^{-1}$, i.e. a spatial dilatation by the factor $\sqrt{\ccc{D_0}/\mathcal{D}_{ii}}$ for each direction $i=x,y,z$, one transforms the anisotropic diffusion tensor $\mathcal{D}$ into the isotropic diffusion tensor $\ccc{D_0}\mathcal{I}$. The domain and the gradient are also affected by this transformation and we denote by prime the new quantities. For instance, spheres are transformed in ellipsoids by this transformation. As the gradient is also affected by the matrix $\mathcal{L}^{-1}$, one has $\mathcal{T}^{(m)}{}' = \mathcal{L}^{-1}\mathcal{T}^{(m)}\mathcal{L}^{-1}$. While the new volume is $V' = \det(\mathcal{L}) V$, there is no simple formula for the surface $S'$ and the $\mathcal{S}^{(3)}{}'$ matrix. Applying our results on isotropic diffusivity to this new case, we get for the effective diffusion coefficient in the original system
\begin{equation}
\der(T)=\ccc{D_0}\left(\Tr(\mathcal{S}^{(2)}\mathcal{T}^{(2)})-\eta'\frac{4}{3\sqrt{\pi}}\frac{S'}{V'}\sqrt{\ccc{D_0}T}+O(T)\right)\;,
\end{equation}
where 
\begin{equation}
\eta'=\Tr(\mathcal{S}^{(3)}{}'\mathcal{T}^{(3)}{}')=\Tr(\mathcal{L}^{-1}\mathcal{S}^{(3)}{}'\mathcal{L}^{-1}\mathcal{T}^{(3)})\;.
\end{equation}
}

\cc{
From the above equation we obtain that $\der(T)$ does not depend (to the order $\sqrt{\ccc{D_0}T}$) on the orientation of the gradient sequence with respect to the medium if $\mathcal{T}^{(2)}$ and $\mathcal{T}^{(3)}$ are isotropic. As we mentioned before, the condition of isotropy of the temporal matrix $\mathcal{T}^{(2)}$ is equivalent to the isotropy of the $b$-matrix that is achieved by spherical encoding techniques \cite{Mori1995a,Cheng1999a,Wong1995a,Graaf2001a,Valette2012a,Eriksson2013a,Topgaard2013a}.
}

\subsection{Multiple compartments}
\label{section:extensions_multiples}

\cc{
Our results were derived under the assumption of a spatially homogeneous intrinsic diffusivity. Moreover, except in Sec. \ref{section:random_medium} where we investigated the effect of orientation dispersion of the confining pores, we implicitly assumed that all confining pores are identical. 
Here we present an extension to a medium that is composed of two or more non-communicating (isolated) compartments (for example, intra- and extra-cellular spaces) with different diffusion coefficients and/or different confining pores .}

\cc{
Inside each compartment, the diffusivity is constant and the pore shapes are identical, so that our formula \eqref{eq:Mitra_gen} for $\der(T)$ is valid, with parameters $\ccc{D_0}$, $\eta$, $S/V$ that depend on the compartment.
The signal can be computed as a voxel-average of signals from individual compartments, and in the regime of small $b$-values ($b\ccc{D_0} \ll 1$), one has, in analogy to Eq. \eqref{eq:signal_average},
\begin{equation}
E\approx\langle \exp(-b\D(T))\rangle \approx \exp(-b\langle \der(T) \rangle)\;,
\end{equation}
where the average is weighted by the relative volume of each compartment, and
\begin{equation}
\langle \der(T) \rangle = \langle \ccc{D_0} \rangle-\frac{4}{3\sqrt{\pi}}\left\langle  \eta\frac{S}{V} \ccc{D_0}^{3/2} \right\rangle \sqrt{T}+O(T)\;.
\end{equation}
We keep this general form of the voxel average which depends on the specific configuration of compartments, pore shapes, diffusivities, etc.

In the above reasoning, the hypothesis of non-communicating compartments is crucial and further modifications would be needed in order to include exchange between compartments when a nucleus can experience different diffusion coefficients during the measurement.
}

\section{Conclusion}
\label{section:discussion}

\color{black}
We presented a generalization of Mitra's formula that is  applicable to any gradient waveform and any geometrical structure. This 
generalized formula differs from the classical one by a \cc{correction} factor in front of $S/V$. In the case of linear encoding schemes, we 
showed that this factor can significantly affect the estimation of $S/V$ and lead to overestimated size of compartments.

We also discussed in detail the effect of anisotropy of  the medium and the use of spherical encoding schemes. In particular, we showed that 
in order to estimate the surface-to-volume ratio of \cc{a mesoscopically} anisotropic medium, the gradient should satisfy the isotropy condition 
($\cc{\mathcal{T}}^{(3)} \propto \mathcal{I}$) that is different from the usual one ($\cc{\mathcal{T}}^{(2)} \propto \mathcal{I}$). In particular, typical spherical 
encoding schemes do not satisfy this new condition. We presented a simple and flexible algorithm that allows fast optimization of gradient 
waveforms and is well-suited for design of diffusion weighted sequences with specific features such as isotropy of $\cc{\mathcal{T}}^{(3)}$, flow compensation, heat limitation, and others.

The developed extension of Mitra's formula \cc{is expected to have} a significant practical impact \cc{due to temporal diffusion encoding parametrization \cite{Lemberskiy2017a,Vellmer2017a}, in particular, in medical applications \cite{Latour1994a,Xu2011a,Reynaud2017a}}.  The proposed approach characterizes the underlying microstructure via novel quantitative metrics such as $\cc{\mathcal{S}^{(3)}}$-tensor and more accurate surface-to-volume ratio. The quantitative scalar maps based on those metrics possess a high potential as a novel set of biomarkers and allow one to apply both well-known diffusion tensor formalism and further improvement of diffusion models based on compartmentization. The practical advantages of the developed approach for designing new gradient encoding schemes for {\em in vivo} brain imaging on clinical scanners will be demonstrated in a separate publication.



\appendices

\section{Theoretical computations}
\label{section:computations}

The signal is proportional to the expectation of the transverse
magnetization which has a form of the characteristic function of the
random dephasing $\phi$ acquired by diffusing spin-carrying molecules:
\begin{equation}
E = \E \{ e^{\cc{-} i \phi} \} , \quad \phi = \gamma \int\limits_0^T B(\vecr(t),t)\der{t}\;,
\end{equation}
where $T$ is the echo time, $\vecr(t)$ is the random trajectory of the
nucleus, $\gamma$ is the gyromagnetic ratio, and $\gamma B(\vecr,t)$ is
the Larmor frequency corresponding to the magnetic field.  In
this work, we consider the most general form of the linear gradient
$\g(t)$:
\begin{equation}
B(\vecr,t) = \g(t) \cdot \vecr = g_x(t) x + g_y(t) y + g_z(t) z.
\end{equation}
In particular, the dephasing can be decomposed as
\begin{equation}
\phi = \phi_x + \phi_y + \phi_z , \quad 
\phi_i = \gamma \int\limits_0^T dt \, g_i(t) (\e_i\cdot \vecr(t))  \quad (i=x,y,z),
\end{equation}
where $\e_x$, $\e_y$ and $\e_z$ are the units vectors in three
directions, and $(\e_i\cdot \vecr(t))$ is the projection of the molecule
position at time $t$ onto the direction $\e_i$.  

The effective diffusion coefficient is related to the second
moment of the dephasing, i.e., we need to evaluate
\begin{equation}
\E\{\phi^2\} = \sum\limits_{i,j=x,y,z} \E\{\phi_i \phi_j\} .
\end{equation}
We emphasize that the three components $\phi_x$, $\phi_y$ and $\phi_z$
are independent only for free diffusion, whereas confinement would
typically make them correlated.  In other words, one cannot {\it a
priori} ignore the cross terms such as $\E\{\phi_x \phi_y\}$.

In order to compute these terms, we use the following representation \cite{Grebenkov2007a}:
\begin{align}  \nonumber
 \E\{\phi_i& \phi_j\} = \gamma^2 \int_0^T \!\!\der{t_1} \int_{t_1}^T \!\der{t_2} \int_\Omega\! \der{\vecr_0} 
\int_\Omega \!\der{\vecr_1} \int_\Omega \!\der{\vecr_2} \int_\Omega \!\der{\vecr_3} \\  \nonumber
& \times \rho(\vecr_0) P_{t_1}(\vecr_0,\vecr_1)  P_{t_2-t_1}(\vecr_1,\vecr_2)  P_{T-t_2}(\vecr_2,\vecr_3) \\
&\times \left[B_i(\vecr_1,t_1) B_j(\vecr_2,t_2) + B_j(\vecr_1,t_1) B_i(\vecr_2,t_2)\right]
\end{align}
where $P_t(\vecr,\vecr')$ is the propagator in the domain $\Omega$, and
$\rho(\vecr_0)$ is the initial density of particles (the initial
magnetization after the $90^\circ$ rf pulse). If the boundary is fully
reflecting and $\rho(\vecr_0)$ is uniform, then the integrals over $\vecr_0$
and $\vecr_3$ yield $1$, so that
\begin{align}  \nonumber
\E\{\phi_i& \phi_j\} = \frac{\gamma^2}{V}\int_0^T \!\der{t_1} \int_{t_1}^T \!\der{t_2}
\int_\Omega \!\der{\vecr_1} \int_\Omega \!\der{\vecr_2} P_{t_2-t_1}(\vecr_1,\vecr_2) \\
 &\times  \left[B_i(\vecr_1,t_1) B_j(\vecr_2,t_2) + B_j(\vecr_1,t_1) B_i(\vecr_2,t_2)\right],
\end{align}
where $V$ is the volume of the domain.  We get thus
\begin{align} \nonumber
\E\{\phi_i &\phi_j\} = \gamma^2\int_0^T \!\der{t_1} \, g_i(t_1)\int_{t_1}^T \!\der{t_2} \, g_j(t_2) K_{ij}(t_2-t_1) \\
&+  \int_0^T \!\der{t_1} \, g_j(t_1)\int_{t_1}^T \!\der{t_2} \, g_i(t_2) K_{ji}(t_2-t_1), 
\end{align}
where
\begin{equation}
K_{ij}(t) = \cc{\frac{1}{V}}\int_\Omega \int_\Omega \cc{p}_i(\vecr_1) P_{t}(\vecr_1,\vecr_2) \cc{p}_j(\vecr_2) \der{\vecr_1}\der{\vecr_2},
\label{eq:Kij}
\end{equation}
with $\cc{p}_i(\vecr) = (\e_i\cdot \vecr)$. Since $K_{ij}(t) = K_{ji}(t)$ due to
the symmetry of the propagator, we can rewrite the moment as
\begin{equation} 
\E\{\phi_i \phi_j\} = \gamma^2 \int_0^T g_i(t_1)\int_{0}^T g_j(t_2) K_{ij}(|t_2-t_1|)\der{t_1}\der{t_2}.
\end{equation}

We rely on the general short-time expansion for the heat kernels (see
\cite{Davies1989a,Gilkey2003a,Desjardins1994a} and references therein)
\begin{equation}  \label{eq:K_exp}
K_{ij}(t) = \sum\limits_{m\geq 0}  c_m(p_i,p_j)\, (\ccc{D_0}t)^{m/2} ,
\end{equation}
with
\begin{subequations}
\begin{align}
c_0(f,\cc{h}) &=  \frac{1}{V} \int_\Omega f(\vecr) \cc{h}(\vecr) \der{\vecr}\;,\\
c_1(f,\cc{h}) &= 0 \;, \\
c_2(f,\cc{h}) &=  - \frac{1}{V} \int_\Omega \nabla f(\vecr) \cdot \nabla \cc{h}(\vecr)\der{\vecr}\;,\\
c_3(f,\cc{h}) &=  \frac{4}{3\sqrt{\pi}} \frac{1}{V} \int_{\partial \Omega} \frac{\partial f(\vecr)}{\partial n} \, \frac{\partial \cc{h}(\vecr)}{\partial n} \der{S}\;,
\end{align}
\end{subequations}
where $\partial/\partial n = (\n\cdot \nabla)$ is the normal
derivative at the boundary, and $\n$ is the unit normal vector at the
boundary oriented outward the domain.  We note that the expansion
(\ref{eq:K_exp}) is an asymptotic series which has to be truncated.
In our case, we get
\begin{subequations}
\begin{align}
c_0(\cc{p}_i,\cc{p}_j) &=  \frac{1}{V} \int_\Omega (\e_i\cdot \vecr) (\e_j\cdot \vecr)\der{\vecr}\;,\\
c_1(\cc{p}_i,\cc{p}_j) &= 0 \;, \\
c_2(\cc{p}_i,\cc{p}_j) &=  - \delta_{ij} \;,\\
c_3(\cc{p}_i,\cc{p}_j) &=  \frac{4}{3\sqrt{\pi}} \frac{1}{V} \int_{\partial \Omega} (\e_i \cdot \n) (\e_j \cdot \n) \der{S}
\end{align}
\end{subequations}
(in the last integral, the normal vector $\n$ depends on the boundary
point).  Combining these results, we get
\begin{align}   \nonumber
&\E\{\phi_i \phi_j\} = \gamma^2 \int_0^T \!\der{t_1} \, g_i(t_1)\int_{0}^T \!\der{t_2} \, g_j(t_2) \\
& \times \!\left(\! - \delta_{ij} \ccc{D_0}|t_2-t_1| + \frac{4}{3\sqrt{\pi}} \frac{S}{V} \cc{\mathcal{S}^{(3)}_{ij}} (\ccc{D_0}|t_2-t_1|)^{3/2} + \cdots\! \right),
\end{align}
where $S$ is the surface area, the ``structural'' matrix $\cc{\mathcal{S}^{(3)}}$ is defined by
%
\begin{equation}
\cc{\mathcal{S}^{(3)}}=\frac{1}{S} \int_{\partial \Omega} \n\otimes\n \der{S}\;,
\end{equation}
and the zeroth order term (with $c_0$) vanished due to the rephasing condition
\begin{equation}  \label{eq:rephasing}
\int_0^T g_i(t)\der{t} = 0 \qquad (i=x,y,z).
\end{equation}
We can write this result more compactly as
\begin{equation}   
\E\{\phi_i \phi_j/2\} = b\ccc{D_0}\left(\delta_{ij} \cc{\mathcal{T}}_{ij}^{(2)} - \frac{4 (\ccc{D_0}T)^{1/2}}{3\sqrt{\pi}} \frac{S}{V} \cc{\mathcal{S}^{(3)}_{ij}}  \cc{\mathcal{T}}_{ij}^{(3)} + \cdots\right) ,
\end{equation}
where we introduced the ``temporal'' matrices
\begin{equation}  \label{eq:Fijn}
\cc{\mathcal{T}^{(m)} = - \frac{\gamma^2T}{2b} \int_0^T \!\!\int_0^T \g(t_1)\otimes\g(t_2)\, \left\lvert\frac{t_2-t_1}{T}\right\rvert^{m/2} \der{t_1}\der{t_2}\; .}
\end{equation}
As a consequence, we compute the second moment as
\begin{equation} \label{eq:phi2_bis}
\cc{\frac{\E\{\phi^2/2\}}{b\cc{D_0}} = \Tr(\cc{\mathcal{T}}^{(2)}) - \frac{4 (\ccc{D_0}T)^{1/2}}{3\sqrt{\pi}} \frac{S}{V} \, \Tr(\cc{\mathcal{S}^{(3)}}\cc{\mathcal{T}}^{(3)}) + \cdots \;.}
\end{equation}

\ccc{Note that this formula can also be obtained from the results of Fr{\o}lich \textit{et al}  \cite{Froehlich2008a}. They compute the effective diffusion coefficient from the velocity auto-correlation function that is then expressed in terms of a double-surface integral of the diffusion propagator. By performing two integration by parts, this integral is essentially identical to our Eq. \eqref{eq:Kij}. The first-order approximation \eqref{eq:phi2_bis} can then be deduced by locally approximating the boundary by a flat surface and using the method of images.}

Let us introduce the auxiliary function
\begin{equation}
\h(t_1)=\int_0^T \g(t_2) \lvert t_2-t_1\rvert\der{t_2}\;.
\label{eq:int_by_parts_start}
\end{equation}
We split the above integral and perform an integration by parts
\begin{align*}
\h(t_1)
&=\frac{1}{\gamma}\int_0^{t_1} \q(t_2) \der{t_2}-\frac{1}{\gamma}\int_{t_1}^T \q(t_2)\der{t_2}\;,
\end{align*}
where we used the conditions $\q(0)=\0$ and $\q(T)=\0$, \cc{with $\q(t)$ being defined in Eq. \eqref{eq:q_def}}. Now we note that
\begin{align*}
\int_0^T &\!\!\!\der{t_1}\int_0^{t_1}\! \g(t_1)\otimes \q(t_2) \der{t_2}=\int_0^T\!\!\!\der{t_2}\int_{t_2}^T \!\g(t_1)\otimes \q(t_2) \der{t_1}\\
&=-\frac{1}{\gamma}\int_0^T \q(t_2)\otimes \q(t_2) \der{t_2}\;,
\end{align*}
where we used again $\q(T)=\0$. In the same way one gets
\begin{equation*}
\int_0^T \!\!\!\der{t_1}\int_{t_1}^T\! \g(t_1)\otimes \q(t_2) \der{t_2}=\frac{1}{\gamma}\int_0^T \q(t_2)\otimes \q(t_2)\der{t_2}\;.
\end{equation*}
Putting all the pieces together, one finally obtains
\begin{align}
\cc{\mathcal{T}}^{(2)}
&=\frac{1}{b}\int_0^T \q(t)\otimes \q(t)\der{t}\;,
\end{align}
so that $\cc{\mathcal{T}}^{(2)}$ is actually the $b$-matrix \cc{renormalized by the $b$-value} \cite{Basser1994a,Mattiello1994a,Basser1994b}. \cc{Since}
\begin{equation}
\Tr(\cc{\mathcal{T}}^{(2)})=\frac{1}{b}\int_0^T \lvert\q(t)\rvert^2\der{t}=1\;,
\label{eq:int_by_parts_end}
\end{equation}
we recover the signal attenuation for free diffusion $E = e^{-\E\{\phi^2/2\}} = e^{-b\ccc{D_0}}$ in the absence of confinement.
In turn, the effective diffusion coefficient, which is experimentally determined from the dependence of $-\ln E$ on $b$ at small b-value, is expressed through the second moment as
\begin{equation} 
\der(T) = \lim\limits_{b\to 0} \frac{-\ln E}{b} = \lim\limits_{b\to 0} \frac{\E\{ \phi^2/2\}}{b}  \;,
\end{equation}
from which, using \eqref{eq:phi2_bis} we obtain Eq. \eqref{eq:Mitra_gen}.

\section{Computation of $\mathcal{S}^{(3)}$ for a weakly perturbed sphere and a spheroid.}
\label{section:computation_S3}
\cc{In this appendix we show an approximate computation of the surface area $S$ and the $\cc{\mathcal{S}^{(3)}}$ matrix of a domain that is a small perturbation of a sphere. Then we provide an exact computation for a spheroid (i.e., an ellipsoid of revolution).}

\subsection{Approximate computation}
\cc{
Let us write the equation of the surface of the domain $\Omega$ in spherical coordinates: $r(\theta,\phi)$, where $r$ is the radius, $\theta$ is the colatitude and $\phi$ the longitude along the surface. We recall that with these conventions, we have an orthogonal basis $(\e_r,\e_\theta,\e_\phi)$, where $\e_r$ is the outward unit radial vector, $\e_\theta$ is directed South along the meridian, and $\e_\phi$ is directed East, perpendicular to $\e_r$ and $\e_\theta$. We also introduce the spherical gradient:
\begin{equation}
\nabla_{\rm s} f= \frac{1}{r}\frac{\partial f}{\partial \theta}\e_\theta + \frac{1}{r\sin\theta}\frac{\partial f}{\partial \phi}\e_\phi\;,
\end{equation}
for a function $f(\theta,\phi)$.}

\cc{
We now write $r(\theta,\phi)=R(1+\varepsilon(\theta,\phi))$, where $\varepsilon(\theta,\phi)$ is a small perturbation. The surface element can then be expressed as
\begin{align}
\der{S} &= r^2 (1+\lVert \nabla_{\rm s} r \rVert^2)^{1/2} \sin{\theta}\der{\theta}\der{\phi}\nonumber\\
&=R^2 (1+2\varepsilon(\theta,\phi))\sin\theta \der{\theta} \der{\phi} + O(\varepsilon^2)\;.
\end{align}
In the same way, one computes the outward normal vector as
\begin{align}
\n &= (1+\lVert \nabla_{\rm s} r \rVert^2)^{-1/2}\left(\e_r-\nabla_{\rm s}r \right)\nonumber\\
&=\e_r-\nabla_{\rm s}r + O(\varepsilon^2)\;.
\end{align}
}

\cc{
Then the surface area of the domain $\Omega$ can be approximated as
\begin{equation}
S\approx 4\pi R^2\left(1 + \frac{1}{2\pi}\int_0^\pi \!\der{\theta} \int_0^{2\pi} \!\der{\phi}\, \varepsilon(\theta,\phi)\sin{\theta}\right)\;.
\end{equation}
In the special case of a domain with a symmetry of revolution, we choose the axis of revolution as the polar axis of the spherical coordinates and get the simpler formula
\begin{equation}
S\approx 4\pi R^2 \left(1+\int_0^\pi \varepsilon(\theta)\sin{\theta}\der{\theta}\right)\;.
\label{eq:S_perturb}
\end{equation}
}

\cc{
Now we turn to the $\cc{\mathcal{S}^{(3)}}$ matrix. As we already obtained $S$, what remains to compute is the following matrix
\begin{equation}
\widetilde{\cc{\mathcal{S}^{(3)}}}=\int_{\partial \Omega} \n\otimes \n \der{S}\;,
\end{equation}
and then $\cc{\mathcal{S}^{(3)}}=\widetilde{\cc{\mathcal{S}^{(3)}}}/S$.
In order to compute the $\widetilde{\cc{\mathcal{S}^{(3)}}}$ matrix, we choose a fixed basis $(\e_x,\e_y,\e_z)$, where $\e_z$ is directed along the polar axis, $\e_x$ corresponds to the direction $\phi=0$ and $\e_y$ to the direction $\phi=\pi/2$. We also introduce the vector $\e_{\rho}$, which is the normalized projection of $\e_r$ on the equatorial plane. In other words, $\e_\rho = \cos(\phi)\e_x+\sin(\phi) \e_y$. Furthermore, we assume that $\Omega$ has a symmetry of revolution around $\e_z$. Thus $\varepsilon$ only depends on $\theta$ and we denote derivative by a prime: $\varepsilon'(\theta)=\frac{\partial \varepsilon}{\partial \theta}$. First we compute the following integral over $\phi$:
\begin{equation}
I(\theta)=\frac{1}{2\pi}\int_0^{2\pi}\left(\e_r-\varepsilon'(\theta) \e_\theta\right)\otimes\left(\e_r-\varepsilon'(\theta) \e_\theta\right)\der{\phi}\;.
\end{equation}
Writing
\begin{subequations}
\begin{equation}
\e_r=\cos(\theta)\e_z+\sin(\theta)\e_\rho\;,
\end{equation}
\begin{equation}
\e_\theta=\cos(\theta)\e_\rho-\sin(\theta)\e_z\;,
\end{equation}
\end{subequations}
we compute:
\begin{subequations}
\begin{equation}
\frac{1}{2\pi} \int_0^{2\pi} \e_\rho \der{\phi} = \0\;,
\end{equation}
\begin{equation}
\frac{1}{2\pi} \int_0^{2\pi} \e_\rho\otimes\e_\rho \der{\phi} = \frac{1}{2} (\e_x\otimes\e_x+\e_y\otimes\e_y)\;.
\end{equation}
\end{subequations}
From the above relations we get 
\begin{align}
I(\theta)&\approx\left(\cos^2\!\theta+\sin(2\theta)\varepsilon'(\theta)\right)\e_z\otimes\e_z\\&+\frac{1}{2}\left(\sin^2\!\theta-\sin(2\theta)\varepsilon'(\theta)\right)(\e_x\otimes\e_x+\e_y\otimes\e_y)\nonumber\;.
\end{align}
The $\widetilde{\cc{\mathcal{S}^{(3)}}}$ matrix is then computed from
\begin{equation}
\widetilde{\cc{\mathcal{S}^{(3)}}}=2\pi\int_0^\pi r^2(\theta) I(\theta)\sin\theta\der{\theta}\;,
\end{equation}
which yields (up to $O(\varepsilon^2)$)
\begin{subequations}
\begin{equation}
\frac{\widetilde{\cc{\mathcal{S}^{(3)}}}_{xx}}{4\pi R^2}=\frac{1}{3}+\frac{1}{2}\int_0^\pi (\varepsilon \sin^3\!\theta- \varepsilon'(\theta) \sin^2\!\theta\cos\theta)\der{\theta}\;,
\end{equation}
\begin{equation}
\widetilde{\cc{\mathcal{S}^{(3)}}}_{yy}=\widetilde{\cc{\mathcal{S}^{(3)}}}_{xx}\;,
\end{equation}
\begin{equation}
\frac{\widetilde{\cc{\mathcal{S}^{(3)}}}_{zz}}{4\pi R^2}=\frac{1}{3}+\int_0^\pi (\varepsilon\cos^2\!\theta\sin\theta+\varepsilon'(\theta) \sin^2\!\theta\cos\theta)\der{\theta}\;,
\end{equation}
\end{subequations}
and the off-diagonal terms are null. Integrating the second terms by part and using \eqref{eq:S_perturb}, we finally get:
\begin{subequations}
\begin{equation}
\mathcal{S}^{(3)}_{xx}=\frac{1}{3}+\int_0^\pi \varepsilon(\theta)\left(\cos^2\!\theta-1/3\right)\sin\theta\der{\theta}+O(\varepsilon^2)\;,
\end{equation}
\begin{equation}
\mathcal{S}^{(3)}_{yy}=\mathcal{S}^{(3)}_{xx}\;,
\end{equation}
\begin{equation}
\mathcal{S}^{(3)}_{zz}=1-2\mathcal{S}^{(3)}_{xx}\;.
\end{equation}
\end{subequations}
}
\cc{
In the case of linear gradient encoding with the gradient oriented either along $\e_x$ or along $\e_z$, the relative variation of $\eta$ is given by (see Eq. \eqref{eq:eta_linear})
\begin{equation}
\frac{\mathcal{S}^{(3)}_{xx}-\mathcal{S}^{(3)}_{zz}}{\mathcal{S}^{(3)}_{zz}}\approx 9\int_0^\pi \varepsilon(\theta)\left(\cos^2\!\theta - 1/3\right)\sin\theta \der{\theta}\;.
\end{equation}
}

\subsection{Exact computation for a spheroid}

\begin{figure}[htp]
\centering
\includegraphics[width=0.3\linewidth]{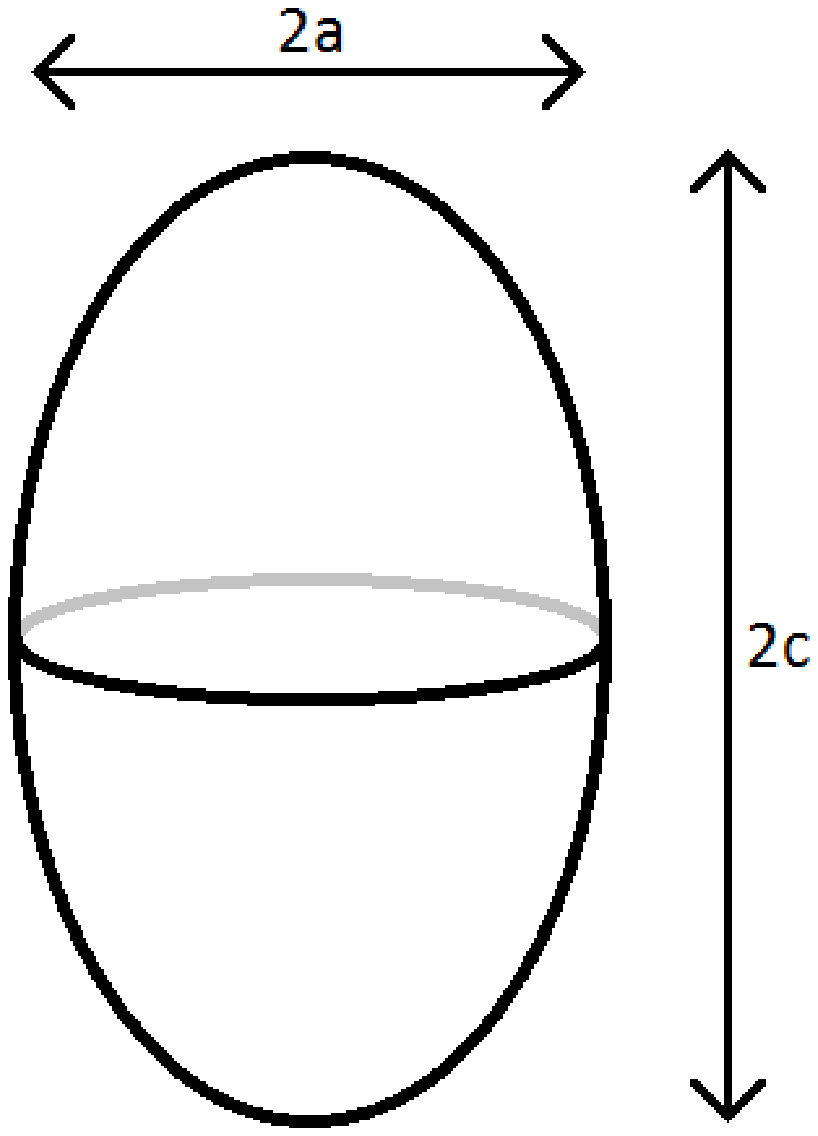}
\includegraphics[height=0.3\linewidth]{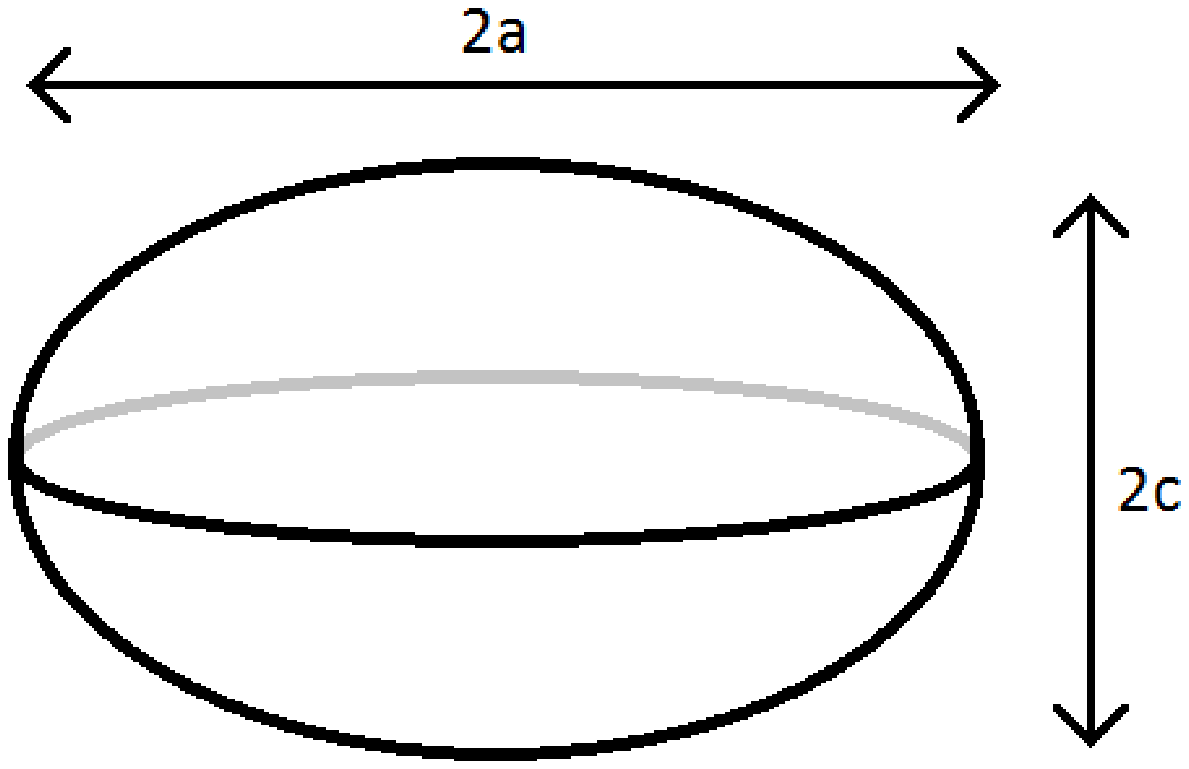}
\caption{\cc{A spheroid (ellipsoid of revolution) is defined by two semi-axes: its equatorial radius $a$ and the distance from the center to the poles $c$. Two situations can occur: (left) the prolate spheroid, with $a \leq c$; (right) the oblate spheroid, with $c\leq a$.}}
\label{fig:ellipsoid}
\end{figure}

\cc{
Let us consider a spheroid (ellipsoid with a symmetry of revolution) with axis $\e_z$. Here we do not consider a small perturbation from a sphere anymore, so that we switch to cylindrical coordinates $(\rho,\phi,z)$ that are more convenient for this computation. Let us recall that $\rho$ is the distance to the revolution axis. The vectors of the basis $(\e_\rho,\e_\phi,\e_z)$ have all been defined in the previous section. We denote by $a$ the equatorial radius of the spheroid and by $c$ the distance from the center to the poles (see Fig. \ref{fig:ellipsoid}). In other words, $a$ and $c$ are the two semi-axes of the spheroid. Two cases will be treated separately: the prolate spheroid ($a\leq c$) and the oblate spheroid $(c\leq a)$. More precisely, we detail the computations for the prolate case and only give the results for the oblate case, as the computations are very similar.
}

\cc{
For the prolate spheroid, we introduce the eccentricity $e$ as $e=\sqrt{1-(a/c)^2}$. Note that $e=0$ corresponds to a sphere of radius $a=c$ and $e=1$ to a stick of length $2c$, oriented along $\e_z$. We have
\begin{equation}
\rho(z)=a\sqrt{1-(z/c)^2}\;,
\end{equation}}
\cc{
and the surface area of the spheroid is readily computed from
\begin{align}
S^{\rm prol}&=2\pi\int_{-c}^{c} \rho(z)\sqrt{1+\left[\rho'(z)\right]^2}\der{z}\nonumber\\
&=2\pi a c \int_{-1}^{1} \sqrt{1-e^2 x^2}\der{x}\;,
\end{align}
which yields
\begin{equation}
S^{\rm prol}=2\pi a c \left(\frac{\arcsin(e)}{e}+\sqrt{1-e^2}\right)\;.
\label{eq:S_prolate}
\end{equation}
For an oblate spheroid, the eccentricity is defined as $e=\sqrt{1-(c/a)^2}$ and the formula for the surface area becomes
\begin{equation}
S^{\rm obl}=2\pi\left(a^2 + c^2 \frac{\artanh(e)}{e}\right)\;.
\label{eq:S_oblate}
\end{equation}
}

\cc{
Now we turn to the computation of $\widetilde{\cc{\mathcal{S}^{(3)}}}$. The outward normal vector is given by
\begin{equation}
(1+\left[\rho'(z)\right]^2)^{-1/2}(\e_{\rho}+\rho'(z)\e_z)\;.
\end{equation}
First we compute the integral over $\phi$:
\begin{align}
I(z)&=\frac{1}{2\pi}\int_0^{2\pi} (\e_{\rho}+\rho'(z)\e_z)\otimes (\e_{\rho}+\rho'(z)\e_z) \der{\phi}\nonumber\\
&=\frac{1}{2} \e_x\otimes\e_x + \frac{1}{2} \e_y\otimes \e_y + \left[\rho'(z)\right]^2 \e_z\otimes \e_z\;.
\end{align}
The $\widetilde{\cc{\mathcal{S}^{(3)}}}$ matrix is then given by
\begin{equation}
\widetilde{\cc{\mathcal{S}^{(3)}}}=2\pi\int_{-c}^{c}\rho(z)(1+\rho'(z)^2)^{-1/2} I(z) \der{z}\;.
\end{equation}
}

\cc{
The following computations assume the prolate case.
Thanks to the relations
\begin{equation}
\widetilde{\cc{\mathcal{S}^{(3)}}}^{\rm prol}_{xx}=\widetilde{\cc{\mathcal{S}^{(3)}}}^{\rm prol}_{yy}=(S^{\rm prol}-\widetilde{\cc{\mathcal{S}^{(3)}}}^{\rm prol}_{zz})/2\;,
\end{equation}we only have to compute $\widetilde{\cc{\mathcal{S}^{(3)}}}^{\rm prol}_{xx}$ in order to have the full $\widetilde{\cc{\mathcal{S}^{(3)}}}^{\rm prol}$ matrix. We have
\begin{align}
\widetilde{\cc{\mathcal{S}^{(3)}}}^{\rm prol}_{xx}&=\pi a c \int_{-1}^{1} \frac{1-x^2}{\sqrt{1-e^2x^2}}\der{x}\\
&=2\pi a c\left(\frac{\arcsin{e}}{e}-\frac{1}{2e^2}\left(\frac{\arcsin{e}}{e}-\sqrt{1-e^2}\right)\right)\nonumber\;,
\end{align}
and then deduce
\begin{equation}
\widetilde{\cc{\mathcal{S}^{(3)}}}^{\rm prol}_{zz}=2\pi a c \left(\frac{\arcsin{e}}{e}-\sqrt{1-e^2}\right)\frac{1-e^2}{e^2}\;.
\end{equation}
Using \eqref{eq:S_prolate}, we come to the matrix $\mathcal{S}^{(3)}$ for the prolate spheroid.
}

\cc{
In the oblate case, one gets
\begin{subequations}
\begin{align}
&\widetilde{\cc{\mathcal{S}^{(3)}}}^{\rm obl}_{xx}=\pi c^2 \left(\frac{\artanh{e}}{e}+\frac{1}{e^2}\left(\frac{\artanh{e}}{e}-1\right)\right)\;,\\
&\widetilde{\cc{\mathcal{S}^{(3)}}}^{\rm obl}_{zz}=2\pi\left(a^2-\frac{c^2}{e^2}\left(\frac{\artanh{e}}{e}-1\right)\right)\;,
\end{align}
\end{subequations}
from which the matrix $\mathcal{S}^{(3)}$ is deduced using \eqref{eq:S_oblate}.
}
\color{black}

\section{\cc{Maximal} value of \cc{$\tau^{(3)}$}}
\label{section:optim}
In the case of linear gradient encoding in a spherical domain, we obtained that \cc{Mitra's formula} is corrected by a factor $\tau^{(3)}$ which is computed from the gradient profile \cc{according to Eq. \eqref{eq:tau_m}.}
In this section, we investigate the maximum and the minimum values of $\cc{\tau^{(3)}}$. 
Integrating by parts (following the same procedure as in Eqs. \eqref{eq:int_by_parts_start}-\eqref{eq:int_by_parts_end}), one obtains
\begin{equation}
\cc{\tau^{(3)}=\frac{3\gamma^2}{8b\ccc{T}}\int_0^T \!\!\int_0^T q(t_1) q(t_2) \left\lvert\frac{t_1-t_2}{T}\right\rvert^{-1/2}\!\der{t_1}\der{t_2}\;.}
\end{equation}
Note that despite its singularity at $0$, the function $1/\sqrt{\lvert t\rvert}$ is integrable, hence the above integral is well-defined.
Next, we apply a change of variables from $t\in[0,T]$ to $t/T \in [0,1]$ and $q(t)$ to $Q(t/T)$, which gives
\begin{equation}
\cc{\tau^{(3)}}=\frac{3}{8 {\lVert Q \rVert}^2}\int_0^1 \int_0^1 Q(t_1) Q(t_2) |t_1-t_2|^{-1/2}\der{t_1}\der{t_2}\;,
\end{equation}
with the usual $L_2$ norm. \cc{One} can understand the above expression as a scalar product
\begin{equation}
\cc{\tau^{(3)}}=\frac{3}{8}\frac{\langle Q, \cc{\mathcal{K}}Q\rangle}{\langle Q,Q\rangle}\;,
\end{equation}
 with an integral operator $\cc{\mathcal{K}}$ with the kernel $\lvert t_1-t_2 \rvert ^{-1/2}$
\begin{equation}
(\cc{\mathcal{K}}Q)(t_1)=\int_0^1 Q(t_2)\lvert t_1-t_2\rvert ^{-1/2}\der{t_2}\;.
\end{equation}
One can see that $\cc{\mathcal{K}}$ is a \cc{weakly singular} convolution operator because the kernel can be expressed  as $K(t_1-t_2)$ (with $K(t)=1/\sqrt{\lvert t 
\rvert}$). Denoting by $\tilde{Q}(\omega)$ the Fourier transform of $Q(t)$ and by $\tilde{K}(\omega)$ the Fourier transform of $K(t)$, one 
gets
\begin{equation}
\langle Q, \cc{\mathcal{K}}Q\rangle
=\frac{1}{2\pi}\int_{-\infty}^{\infty} \lvert \tilde{Q}(\omega)\rvert^2 \tilde{K}(\omega)\der{\omega}\;,
\label{eq:alpha_Fourier}
\end{equation}
with $\tilde{K}(\omega)=\sqrt{2\pi/\lvert \omega \rvert}$. This shows that $\cc{\tau^{(3)}}$ is always positive (in other words, the operator 
$\cc{\mathcal{K}}$ is positive-definite). This result is expected from a physical point of view: if $\cc{\tau^{(3)}}$ were negative, 
then the effective diffusion coefficient would increase with time \cc{that is nonphysical}. The minimum value $0$ can be 
asymptotically obtained, for example, with very fast oscillating gradients. It is, indeed, clear from 
\cc{Eq.} \eqref{eq:alpha_Fourier} that if $g(t)$ is a cosine function with angular frequency $\omega_0$ such that the number of periods $N_0=\omega_0 
T/(2\pi) \gg 1$, then $\lvert \tilde{Q}(\omega)\rvert^2$ is concentrated around $\pm \omega_0$, and we obtain $\cc{\tau^{(3)}}\approx 
3/(8\sqrt{N_0})\cc{\sim \omega_0^{-1/2}}$, \cc{a result that was obtained as well in} \cite{Novikov2011b} (see also Fig. \ref{fig:profils}).

Now we turn to the maximum value of $\cc{\tau^{(3)}}$. The condition that $Q(t)$ is null outside of $[0,1]$ is difficult to take into account  in 
Fourier space and we could not extract further information from Eq. \eqref{eq:alpha_Fourier}. In order to bound the maximum value of 
$\cc{\tau^{(3)}}$, one can use the Cauchy inequality:
\begin{align}
&\lvert(\cc{\mathcal{K}}Q)(t_1)\rvert=\left\lvert\int_0^1 \sqrt{K(t_1-t_2)}Q(t_2)\sqrt{K(t_1-t_2)}\der{t_2}\right\rvert \nonumber\\
&\;\leq \left(\int_0^1 K(t_1-t_2)\der{t_2}\right)^{\frac{1}{2}} \left(\int_0^1 Q^2(t_2)K(t_1-t_2)\der{t_2}\right)^{\frac{1}{2}}\nonumber\;.
\end{align}
One can easily compute the function
\begin{equation}
\int_0^1K(t_1-t_2)\der{t_2}=2\sqrt{t_1}+2\sqrt{1-t_1}\;,
\end{equation}
whose maximum is $2\sqrt{2}$. Thus, one gets
\begin{equation}
\lvert(\cc{\mathcal{K}}Q)(t_1)\rvert\leq \left(2\sqrt{2}\int_0^1 Q^2(t_2)K(t_1-t_2)\der{t_2}\right)^{\frac{1}{2}}\;.
\end{equation}
Using again the Cauchy inequality, one obtains
\begin{equation*}
\langle Q,\cc{\mathcal{K}}Q\rangle \leq 2^{3/4}\lVert Q\rVert \left(\int_0^1\int_0^1 Q^2(t_2)K(t_1-t_2)\der{t_2}\der{t_1}\right)^{\frac{1}{2}}\;.
\end{equation*}
The same reasoning about the maximum value of the integral of $K$
yields
\begin{equation}
\langle Q,\cc{\mathcal{K}} Q \rangle \leq 2^{3/2} \lVert Q \rVert^2\;,
\end{equation}
and finally
\begin{equation}
\cc{\tau^{(3)}} \leq \frac{3\sqrt{2}}{4} \approx 1.06\;.
\end{equation}
We also know from the examples in Fig. \ref{fig:profils} that $\cc{\tau^{(3)}}=1$ can be achieved \cc{for $Q\equiv 1$}, which implies that the maximum value of 
$\cc{\tau^{(3)}}$ is in the interval $[1,1.06]$.

The problem can be considered from another point of view. Due to the symmetry of the operator $\cc{\mathcal{K}}$, it is 
well-known that the function $Q$ maximizing $\langle Q,\cc{\mathcal{K}}Q\rangle/\lVert Q \rVert^2$ is the eigenfunction of $\cc{\mathcal{K}}$ 
with the highest eigenvalue. As a consequence, if one searches for a good estimation of the maximum $\cc{\tau^{(3)}}$ as well as the 
corresponding ``optimal'' gradient profile, then one can use the following procedure: (i) to choose an initial profile $Q_0$ which is 
sufficiently general or sufficiently close to a guessed optimal profile; (ii) to apply iteratively the operator $\cc{\mathcal{K}}$ and to 
renormalize the result; (iii) to stop when the sequence has converged.

For example, the initial profile $Q_0(t/T)=1$, which corresponds to two infinitely narrow gradient pulses  at time $0$ and $T$, yields 
$\cc{\tau^{(3)}}=1$, which is close to the optimal value. Thus, it is a good initial condition for the iterative process.
\begin{figure}[htbp]
\begin{center}
\includegraphics[width=0.9\linewidth]{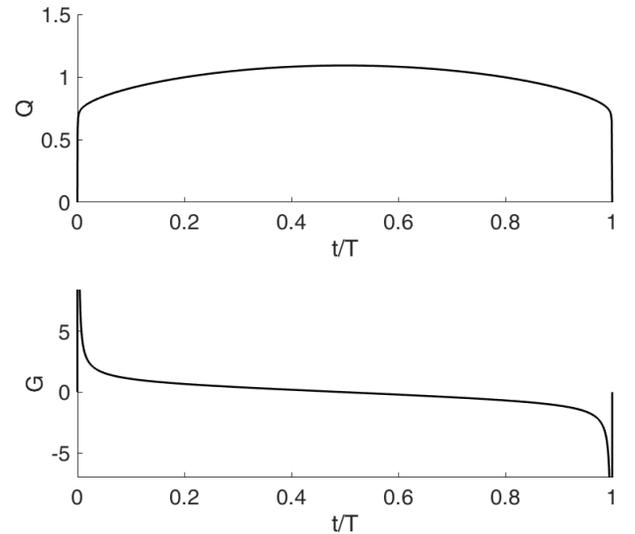}
\caption{The result of the iterative procedure in order to obtain the optimal profile that maximizes the value of $\cc{\tau^{(3)}}$.}
\label{fig:optim}
\end{center}
\end{figure}
The result of such a procedure is shown in Fig. \ref{fig:optim}.  This yields an optimum value of $\cc{\tau^{(3)}}$ of about $1.006$, thus very close 
to $1$. It is worth to note, however, that the optimal profile $Q(t/T)$ differs clearly from $Q_0(t/T)=1$ \cc{(note also that $Q_0$ is not an eigenfunction of $\cc{\mathcal{K}}$)}.

\color{black}

\section{Fully isotropic sequence}
\label{section:full_iso}

\cc{
The conventional condition $\mathcal{T}^{(2)} \propto \mathcal{I}$ removes the microscopic anisotropy in the diffusion tensor, whereas the new isotropy condition $\mathcal{T}^{(3)} \propto \mathcal{I}$ eliminates the mesoscopic anisotropy in the leading order of the short-time expansion. One can thus naturally ask whether it is possible to design a ``fully isotropic'' sequence that removes anisotropy in all order of $(\ccc{D_0}T)^{1/2}$?
}
\cc{
In this appendix we show that it is impossible to find a gradient sequence such that $\mathcal{T}^{(m)}$ is isotropic for all integer values $m=2,3,4,\ldots$. In other words, one cannot find a sequence which produces an isotropic time-dependence of $\der(T)$ to every order in $(\ccc{D_0}T)^{1/2}$.
To show this we restrict ourselves to the values of $m$ that are multiple of $4$, $m=4l$, with $l=1,2,\ldots$. 
\begin{align}
\mathcal{T}^{(4l)}_{ij}&=-\frac{\gamma^2 T^{1-2l}}{2b}\int_0^T\!\!\int_0^T g_i(t_1)g_j(t_2)(t_2-t_1)^{2l}\der{t_1}\!\der{t_2}\nonumber\\
&=-\frac{\gamma^2 T^{1-2l}}{2b} \sum_{k=0}^{2l} (-1)^k \binom{2l}{k}\alpha_i^{(k)}\alpha_j^{(2l-k)}\;,
\end{align}
where 
\begin{equation}
\alpha_i^{(k)}=\int_0^T g_i(t) t^k \der{t}\;.
\end{equation}
}

\cc{
We will now prove that the isotropy of $\mathcal{T}^{(4l)}$ for any integer $l$ implies that $\alpha_i^{(k)}=0$ for all $i=x,y,z$ and all integer $k$. Note that the property for $k=0$ corresponds to the refocusing condition \eqref{eq:refocusing} that we assumed throughout the paper. We prove our statement by recurrence on $l$ and $k$.
First, let us consider $l=1$ and prove the $k=1$ case. One has
\begin{equation}
\mathcal{T}^{(4)}_{ij}=-\frac{\gamma^2 T^{-1}}{2b}\left(-2\alpha_i^{(1)}\alpha_j^{(1)}\right)\;.
\end{equation}
If $i\neq j$, then $\mathcal{T}^{(4)}_{ii}=\mathcal{T}^{(4)}_{jj}$ and $\mathcal{T}^{(4)}_{ij}=0$ so that $\alpha_i^{(1)}=\alpha_j^{(1)}=0$.
}

\cc{
Now we assume that $\alpha_i^{(k)}=0$ for all $i=x,y,z$ and for all $k<k'$ up to a given rank $k'$. Then almost all the terms in the expression of $\mathcal{T}^{(4k')}_{ij}$ vanish and we are left with
\begin{equation}
\mathcal{T}^{(4k')}_{ij}=-\frac{\gamma^2 T^{1-2k'}}{2b}\left((-1)^{k'}\binom{2k'}{k'} \alpha_i^{(k')}\alpha_j^{(k')}\right)\;,
\end{equation}
and with the same reasoning as in the previous case, we deduce that $\alpha_i^{(k')}=0$ for any $i$. By recurrence, we have proven that $\alpha_i^{(k)}=0$ for all $i$ and $k$.
}

\cc{
What remains to prove is that the only continuous function $f(t)$ that satisfies $\int_0^T f(t)t^k\d{t}=0$ for all integer values of $k$ is the null function $f=0$. Let us assume that $f$ is nonzero, i.e., there exists an interval $(a,b)$ with $a<b$ such that $f(t)\neq 0$ for any $t\in (a,b)$ (e.g., $f(t)>0$ on this interval). Since polynomials form a dense subset of continuous functions on $[0,T]$, one can build a sequence of polynomials that converges to a continuous function that would be zero outside $(a,b)$ and positive inside $(a,b)$. Thus there would exist a polynomial $P(t)$ such that $\int_0^T f(t)P(t)\der{t}>0$, which is incompatible with the statement: for all $k=0,1,2,\ldots$ $\int_0^T f(t)t^k \der{t}=0$. 
Note that this argument can be easily extended to functions with a finite number of jumps.

}
\color{black}

\end{document}